\documentclass[fleqn,usenatbib,useAMS]{mnras}
\usepackage{graphicx}	
\usepackage{amsmath}	
\usepackage{amssymb}	
\usepackage{longtable}

\usepackage{multicol}        
\usepackage{bm}		
\usepackage{pdflscape}	
\usepackage[table,xcdraw]{xcolor}
\definecolor{Blue}{rgb}{0,0.08,0.65}

\newcommand{\mass}{{\mathcal{M}}}
\newcommand{\ssfr}{{\mathcal{S}}}

\newcommand{\gm}[1]{{\color{black}{{#1}}}}
\newcommand{\sk}[1]{{\color{black}{{#1}}}}

\usepackage[T1]{fontenc}
\usepackage{ae,aecompl}

\usepackage{multirow}

\usepackage{txfonts}
\DeclareMathOperator\erf{erf}

\title[The SFRs of Elliptical Galaxies from CCSNe]{The Star Formation Rates of Elliptical Galaxies\\ from Core-Collapse Supernovae}

\author[T. M. Sedgwick et al.]{T. M. Sedgwick$^{1}$\thanks{Contact e-mail: \href{mailto:T.M.Sedgwick@2013.ljmu.ac.uk}{T.M.Sedgwick@2013.ljmu.ac.uk}},
I. K. Baldry$^{1}$, P. A. James$^{1}$, S. Kaviraj$^{2}$ \& G. Martin$^{3,4}$\\
$^{1}$Astrophysics Research Institute, Liverpool John Moores University, IC2, Liverpool Science Park, 146 Brownlow Hill, Liverpool L3 5RF, UK\\
$^{2}$Centre for Astrophysics Research, School of Physics, Astronomy and Mathematics, University of Hertfordshire, Hatfield AL10 9AB, UK\\
$^{3}$Steward Observatory, University of Arizona, 933 N. Cherry Ave, Tucson, AZ 85719, USA\\ 
$^{4}$Korea Astronomy and Space Science Institute, 776 Daedeokdae-ro, Yuseong-gu, Daejeon 34055, Korea}

\date{Submitted to MNRAS, 2021 June.}

\pubyear{2021}

\begin{document}
\label{firstpage}
\pagerange{\pageref{firstpage}--\pageref{lastpage}}
\maketitle

\begin{abstract}

The level of star formation in elliptical galaxies is poorly constrained, due to difficulties in quantifying the contamination of flux-based estimates of star formation from unrelated phenomena, such as AGN and old stellar populations. We here utilise core-collapse supernovae (CCSNe) as unambiguous tracers of recent star formation in ellipticals within a cosmic volume. We firstly isolate a sample of 421 $z < 0.2$, $r < 21.8$ mag CCSNe from the SDSS-II Supernova Survey. We then introduce a Bayesian method of identifying ellipticals via their colours and morphologies in a manner unbiased by redshift and yet consistent with manual classification from Galaxy Zoo 1. We find $\sim 25\%$ of $z < 0.2$ $r < 20$ mag galaxies in the Stripe 82 region are ellipticals ($\sim 28000$ galaxies). In total, 36 CCSNe are found to reside in ellipticals. We demonstrate that such early-types contribute a non-negligible fraction of star formation to the present-day cosmic budget, at {$11.2 \pm 3.1$ (stat) $^{+3.0}_{-4.2}$ (sys) \%}. Coupling this result with the galaxy stellar mass function of ellipticals, the mean specific star formation rate (SSFR; $\overline{\mathcal{S}}$) of these systems is derived. The best-fit slope is given by $\log (\overline{\mathcal{S}}(\mathcal{M})/\rm{yr}) = -\; (0.80 \;\pm\;0.59) \; \log (\mathcal{M}/10^{10.5}\rm{M}_{\odot}) \; - 10.83 \;\pm\;0.18$. The mean SSFR for all {$\log (\mathcal{M}/\rm{M}_{\odot}) > 10.0$} ellipticals is found to be {$\overline{\mathcal{S}} = 9.2 \pm 2.4$ (stat) $^{+2.7}_{-2.3}$ (sys) $\times 10^{-12}$ yr$^{-1}$}, which is consistent with recent estimates via SED-fitting, and is {$11.8 \pm 3.7$ (stat) $^{+3.5}_{-2.9}$ (sys) \%} of the mean SSFR level on the main sequence as also derived from CCSNe. We find the median optical spectrum of elliptical CCSN hosts is statistically consistent with that of a control sample of ellipticals that do not host CCSNe, implying that these SN-derived results are well-representative of the total low-$z$ elliptical population.


\end{abstract}


\section{Introduction}

Elliptical galaxies dominate the high-mass end of the stellar mass function at late epochs \citep[e.g.][]{KEL14b}, making them essential probes of our structure formation paradigm. {The stellar mass growth of these systems is tightly linked to their star formation rates over cosmic time}. The classical hypothesis of a `monolithic collapse' \citep{ELBS62,LAR74,CC02} suggests that the stellar populations of early-type galaxies formed at high redshifts ($z \gtrsim 2$), and that these galaxies have since evolved passively.

Optical properties of ellipticals appear to obey simple scaling relations which indicate that the majority of star formation in ellipticals galaxies did indeed occur at high redshift. For instance, both rest-frame optical colours and the `Fundamental Plane' of ellipticals \citep{JFK96,SAG97} exhibit only a small scatter, and do not appear to have evolved significantly for several Gyrs \citep[e.g.][]{BLE92,FRA93,VD00,PEE02}.

Problematically, a monolithic star formation history cannot be aligned with the currently widely-accepted $\Lambda$-CDM framework, which in both semi-analytical models \citep[e.g.][]{COL00,KB03} and in recent hydrodynamical simulations \citep[e.g.][]{SCH15,PIL18} predicts ongoing contributions to early-type galaxy stellar mass from merger and interaction driven star formation \citep{MAR17}. Therefore, an accurate quantification of the star formation levels in ellipticals has ramifications for our knowledge of both early-type galaxy evolution and cosmology.

The star-formation rates of early-type galaxies can be estimated observationally via several means. UV emission is a widely utilised indicator, as it exhibits the signal of photospheric light from the very youngest, most massive stars \citep[e.g.][]{SAL07,BOU09}. A less direct indicator is mid and far IR emission, which captures the heating of dust due to young stellar populations \citep[e.g.][]{CAL00,INO00}. Specific nebular recombination lines are also widely utilised, most notably, {H$\alpha$}, which has contributions to its flux from hot, luminous, young stars \citep[e.g.][]{GLA99,ERB06}.

However, each of these diagnostics can have their signal contaminated by other mechanisms. For example, post-AGB stars see their outer envelopes, in the process of being ejected, illuminated by core UV emission \citep{GRE90}. In Horizontal Branch stars, the intense temperatures of core-helium burning can also cause UV emission in these stars' optically thin regions. In the case of both phenomena, strong emission lines such as {H$\alpha$} can also be generated; lines which are typically associated with the presence of young stars \citep{PRE06}. Other mechanisms which could be misinterpreted as star formation without careful investigation include the ionisation of gas due to active galactic nuclei \citep[AGN;][]{GRO06}, or due to shocks, with the collision of gas clouds \citep{OPA18}.

In the case of galaxies residing on the star forming main sequence \citep{NOE07}, such as most spiral galaxies, the contributions from these contaminants make up a small fraction of the star formation signal, and can be largely ignored. However, for galaxies below the main sequence, such as ellipticals, these contaminants present a serious boundary to studies of their star formation rates. 

\sk{A consequence of this issue is some degree of debate over the precise level of star formation typical in ellipticals, which in turn perpetuates the debate over our pictures of early-type galaxy mass assembly and cosmology. Whilst some studies conclude practically zero levels, \citep[e.g.][]{JUR77,KEN98},

more recent work suggests that low-level star formation persists in ellipticals over at least the latter half of cosmic time. For instance, \citet{KAV07} find that, while they are a homogeneous population in optical colours \citep{BOW92}, elliptical galaxies show more than 6 mags of colour spread in the NUV, which is a much wider range than can be produced from old stellar populations \citep[e.g.][]{YI03}. They conclude that at least 30\% of local elliptical galaxies have NUV colours indicative of recent star formation within the last Gyr, and at $z \lesssim 0.1$, elliptical galaxies have {$1\% - 3\%$} of their stellar mass in stars produced within such a time-frame.} 

\sk{Interestingly, for ellipticals at intermediate redshift ($0.5<z<1$), i.e. epochs at which the Universe is effectively too young for old stellar populations to exist, the characteristics of NUV colours remain the same \citep{KAV08,MAR18}. This implies that this observed low-level star formation persists in early-type galaxies over at least the latter half of cosmic time. There also exists a strong correspondence between the presence of disturbed morphologies and blue UV colours \citep{KAV11} indicating that the star formation is merger driven \citep{SCH90,SS92}. Furthermore, the frequency of these morphological disturbances is larger than can be accounted for by the major-merger rate alone \citep{KAV11}, suggesting that minor mergers (i.e. mergers between ellipticals and gas-rich satellites) are likely to drive much of the elliptical star formation seen in the NUV, at least at late epochs \citep{KAV09}, with $\sim 14 \%$ of the total cosmic star formation budget residing in the elliptical population in the low-redshift Universe \citep{KAV14a,KAV14b}.}

Broad-wavelength SED-based derivations, such as those from \citet{SAL16}, offer arguably the most sensitive treatment to sub-main sequence star formation rates to date, due to the ability to better identify sources of contamination to the star formation signal from the broad view of spectral information spanning from the UV to the IR. However, even these estimates are only reported as upper limits on the level of star formation in ellipticals.

It is common to try to distinguish signatures of true star formation from those of AGN or old stellar populations using galaxy emission line ratios, such as H$\alpha$/[NII] and $H\beta$/[OIII] \citep[e.g.][]{KAU03}. However, there is scatter in these relations, and if ellipticals do exhibit some level of star formation, they are unlikely to occupy the main locus as blue star-forming galaxies. As such, the issue is turned on its head, and low-level star formation could easily be misattributed to AGN activity and/or old stellar emission in a discrete treatment, such as that seen with the BPT diagram \citep[e.g.][]{BPT81,VO87}. Additionally, there are not an abundance of models which can predict these ratios from first principles. 

Recent sophisticated simulation suites such as EAGLE \citep{SCH15}, Horizon-AGN \citep{DUB16,KAV17} and IllustrisTNG \citep{PIL18} support the suggestion that minor mergers could generate a persistent low-level of star formation. This would not only make them a phenomenon crucial to the evolution of ellipticals, but would also point further towards the hierarchical evolution now widely accepted under a $\Lambda$-CDM paradigm. 

Such simulations also implement sub-grid star formation recipes as well as feedback mechanisms which can remove cold gas and quench star formation. \gm{Since these sub-grid recipes are often calibrated and benchmarked against a number of observed properties and relations \citep[including star formation rate densities; e.g.][]{PIL18}, any biases or systematic uncertainties in these quantities can produce unrealistic sub-grid physics and degrade the predictive power of the simulations.} 

If the observed elliptical morphology and colour are reproduced organically via `dry' mergers as a consequence of quenching through feedback, both merger rates (minor and major) and feedback mechanisms could be better constrained in light of accurate and unbiased tracers of elliptical star formation rates. Furthermore, since many of these simulations now aim to replicate the star formation rate density (SFRD) of main-sequence galaxies as standard, a natural progression would be to test model physics using the observed star formation rates of the early-type population. 

\gm{There are other mechanisms which may also drive star formation in ellipticals. For example, galaxies of all morphological types are surrounded by significant reservoirs of cold gas \citep[][]{Chen2010,Thom2012}, which may fuel low levels of star formation even in apparently dead ellipticals \citep[][]{Tumlinson2017}. In hydrodynamical simulations, the structure and physical properties of the CGM are found to depend strongly on factors such as resolution, feedback and the presence of magnetic fields and cosmic rays \citep[][]{Hummels2019,Butsky2020}. Therefore, accurate measurements of the level of star formation in elliptical galaxies are an important clue for understanding the connection between galaxy evolution and the CGM. In more sophisticated simulations, which incorporate magnetic fields and cosmic ray heating in addition to stellar and AGN feedback processes, observations like these will be important for placing constraints on the recipes used to model these processes.}

For each of the described reasons, we are encouraged to search for an independent diagnostic of star formation which is not subject to the aforementioned uncertainties. Core-collapse supernovae (CCSNe) are unambiguous indicators of recent star formation. The short \citep[$\sim$ 6-40 Myr;][]{SMA09,BOT17} lifetimes of their progenitor stars, relative to the timescales of galaxy evolution, mean that the presence of CCSNe in the low-$z$ elliptical population would offer explicit evidence of recent low-level star formation within them.

These transients also provide a more thorough analysis of star formation in the shape of the volumetric SFRD ($\rho_{\rm{SFR}}$) as a function of galaxy stellar mass, $\mathcal{M}$. It was shown in \citet{SED19b} and \citet{SED19a} that the SFRD of low-surface brightness galaxies can be accurately traced from the volume's CCSN rates, $\rho_{\rm{CCSN}}$, as measured by the untargeted, high-cadence, SDSS-II Supernova Survey \citep{SAK18}, by matching CCSNe to their host galaxies. This connection is shown via Equation~\ref{eq:CCSNSFRD}, where $\overline{\mathcal{R}}$ is the mean ratio of CCSN events to mass of stars formed.

\begin{equation}\label{eq:CCSNSFRD}
  \rho_{\rm SFR}(\mass) \: = \: \frac{\rho_{\rm CCSN}(\mass)}{\overline{\mathcal{R}}}
\end{equation}

In the case of low surface brightness galaxies (LSBGs), CCSNe not only helped locate galaxies which would have been missed using traditional galaxy extraction pipelines, by acting as signposts towards them, but they also allowed the population's star formation to be quantified in situations where spectral and broad-band photometry-based derivations would be subject to large uncertainties related to low signal-to-noise. It is this latter point which particularly encourages us to extend a similar method to elliptical galaxies. 

The SFRD would allow us to place constraints on the elliptical contribution to the cosmic star formation budget. As well as the SFRD, we may also be able to assess the mean \textit{specific} star formation rate (SSFR; $\overline{\mathcal{S}}$) of ellipticals, both as a function of mass, and for the total population. This is due to the connection between the SFRD, the galaxy stellar mass function (GSMF) and $(\overline{\mathcal{S}})$, demonstrated via Equation~\ref{eq:Sbar}, where $\phi$ is the number density of galaxies in a cosmological volume.

\begin{equation}\label{eq:Sbar}
     \overline{\ssfr}(\mass) \: = \: \frac{\rho_{\rm SFR}(\mass)}{\Phi_{\rm}(\mass) \: \mass}
\end{equation}

In \citet{SED19a}, the SFRD was first found from CCSNe, before the GSMF was determined using a re-arrangement of Equation~\ref{eq:Sbar} from the assumption that $\overline{\mathcal{S}}$ is constant with mass for low-surface brightness galaxies, due to their residence on the star forming main sequence \citep{MCG17}. In the case of massive ellipticals, however, the GSMF is already well-constrained via various experiments \citep{FRA06,VUL11,KEL14b}, and it is instead $\overline{\mathcal{S}}$ which is our unknown. Whilst we will here utilise Equation~\ref{eq:Sbar} to measure the mean SSFR of ellipticals, the relation can be applied to any well-defined galaxy population for which any two of $\rho_{\rm{SFR}}$, $\phi$ and $\overline{\mathcal{S}}$ are known.

The structure of the present work is summarised as follows: Sections~\ref{sec:CCSN} and \ref{sec:ell} summarise the definition of the CCSN and elliptical samples, respectively along with summaries of the data sets used. Section~\ref{sec:SFRD} gives measurements of the volumetric star formation rate density of ellipticals. Section~\ref{sec:SSFR} shows results for the galaxy stellar mass function and mean specific star formation rate of ellipticals. Section~\ref{sec:spec} compares the spectra of CCSN-hosting ellipticals to that of a control sample of standard ellipticals. Finally, Section~\ref{sec:summary} summarises each of our findings in the context of early-type galaxy evolution.

\section{The CCSN sample}\label{sec:CCSN}

We begin with the SN sample utilised in \citet{SED19a}. This was formed from the SDSS-II Supernovae Sample \citep{SAK18}, cross matched with the IAC Stripe 82 legacy sample \citep{FT16}. A more detailed discussion of this cross-matching is given in \citet{SED19a}, which describes the careful assignment of host galaxies, and the rejection of variable stars and AGN. This sample was cut to exclude SN of $r$-band peak brightness > 21.8 magnitudes, fainter than which the sample is estimated to be incomplete (as also described in the aforementioned study). This initial sample consists of 2528 $r_{peak}<21.8$ mag SNe and their hosts, following variable star/AGN removal.

We utilise the best redshift estimate available for each SN/host pair: this is spectroscopic, from either SN (taking preference) or host galaxy, the latter from either SDSS-II Legacy \citep{YOR00}, SDSS-II Southern \citep{BAL05}, SDSS-III BOSS \citep{DAW13} or SDSS-IV eBOSS \citep{DAW16}, or in the absence of spectra, photometric, from $ugriz$ Stripe 82 legacy survey galaxy fluxes, using the `scaled flux matching' technique \citep{SED19b,BAL21}. The SN sample is then cut to $z < 0.2$, leaving 1070 SNe, 845 of which have a spectroscopic redshift.

In the present work, we treat SN classification differently than in \citet{SED19b} and \citet{SED19a}. \citet{SAK18} attempted to classify each of these SNe as either a Type Ia, Type Ib/c, or Type II. As a brief summary, 2 main sets of probabilities were calculated for each SN: 

The first was a set of Bayesian probabilities of belonging to each of the 3 SN classes, obtained by analysing observed light curves with the Photometric SN IDentification (PSNID) software \citep{SAK11}, using a grid of Type Ia, Type Ib/c and Type II SN light curve templates.

The second was a set of Bayesian + nearest-neighbour (NN) probabilities, the improvement being that with the NN extension, probabilities account for differences in the distributions of extinction, light curve shape and redshift seen for each of the 3 SN types. Synthetic SNe of each type were considered a `near-neighbour’ to an observed SN if within a threshold Cartesian distance in a 3-dimensional parameter space consisting of the above 3 parameters. The NN probability of the candidate being a Type II SN, for instance, was then the fraction of NN simulated SNe which were Type II's. In the present work, we use the NN probability, $P_{\rm{NN}}$ if there are at least 20 near-neighbours for the transient, which is the case for 738 {$z < 0.2$} SNe, and in the 332 remaining cases we use the Bayesian-only probability, $P_{\rm{Bayes}}$. We denote the best probabilities available as $P_{\rm{Ia}}$, $P_{\rm{Ibc}}$ and $P_{\rm{II}}$. Though there is some spread in the $P_{\rm{NN}} - P_{\rm{Bayes}}$ distributions, with a standard deviation of 0.22 for the $z < 0.2$ Type Ia probabilities, for example, the mean difference of $P_{\rm{NN}} - P_{\rm{Bayes}}$ for Type Ia SNe is very close to zero, implying a consistency between the 2 methods on a statistical level.

Elliptical galaxies are more traditionally associated with Type Ia SNe than with CCSNe \citep[e.g.][]{GAL08}. To ensure Type Ia interlopers do not cause an overestimate of star-formation properties, we therefore require $P_{\rm{Ia}} < 0.05$ for the SN to be included in our CCSN sample. Removing those with $P_{\rm{Ia}} > 0.05$, we are left with 360 confidently classified $z < 0.2$ CCSNe. This sub-sample is used for all main CCSN-based results in the present work. However, we will repeat our analysis including an additional 61 $0.05 < P_{\rm{Ia}} < 0.5$ objects, to test for the sensitivity of results to Type Ia contamination. Note that these numbers suggest the CCSN probabilities are largely bi-modal. Put differently, $\sim 93\%$ of all $z < 0.2$ SNe have either {$P_{\rm{Ia}} < 0.05$} or {$P_{\rm{Ia}} > 0.95$}.

The ratio of likely Type Ib/c SNe to likely Type II SNe is also approximately in line with expectations from the literature \citep[e.g.][]{HAK08}, with $24 \pm 1$ (Poisson error) $\%$ having $P_{\rm{Ibc}} > P_{\rm{II}}$. Furthermore, it is far more common to misclassify a Type Ia as a Type Ic rather than as a Type II SN, due to the similar light curve shapes of Type Ia and Type Ic supernovae (lack of plateau), and due to their similar spectra \citep{CLO97}. As a result, if we had significant contamination from Type Ia supernovae in our core-collapse sample, we would expect a higher ratio of Type Ib/c to Type II supernovae than is observed. These points increase our confidence that the Bayesian SN-type classifications are fully trustworthy.

Other tests for the sensitivity of forthcoming results to contamination include a repeating of results with SN type probabilities re-scaled to account for the efficiency and purity estimates in each SN class. These values were estimated from a set of simulated SNe by \citet{SAK18}. We also test the effect of simply using Bayesian-only probabilities for all SNe. As might be expected given the aforementioned mean of the $P_{\rm{NN}} - P_{\rm{Bayes}}$ distribution, and the bi-modality of the CCSN probabilities, respectively, we find for both tests that our forthcoming results do not change notably. In conclusion, the CCSNe are classified from their light curves with sufficient confidence that our sample selection is stable to the fine tuning of the SN classification procedure.

\section{Defining elliptical galaxies}\label{sec:ell}

We must next determine which hosts galaxies are ellipticals, and as quantitatively as possible. For this we turn to the wider sample of galaxies in Stripe 82 legacy coadded imaging. 

Using estimates by \citet{FT16} from simulated de Vaucouleurs profiles, we can expect $95\%$ of all $z < 0.2$ bulge-dominated galaxies in the legacy coadded imaging to have at least $95\%$ of their light recovered at $r_{\textsc{auto}} = 20$ mag, where $\textsc{auto}$ denotes the near-total light `Kron' elliptical aperture \citep{KRO80,BA96}. We therefore impose this magnitude cut, leaving a sample of 113239 galaxies. Note that we do not apply this cut to the SN hosts, as it is the SN light which determines their detection. Nonetheless, 70\% of all $z < 0.2$ SN hosts in our sample have $r < 20$ mag, and this rises to 98\% for $\mathcal{M} > 10^{10.0} \rm{M}_{\odot}$, masses below which we expect few ellipticals.

The most common and straightforward way to attempt to isolate ellipticals is to use a cut on galaxy morphology. For example, \citet{KEL14b} define ellipticals as all single-component, bulge-dominated galaxies. A useful parameter to quantitatively represent `bulge dominance', available for all SDSS galaxies, is the `de Vaucouleurs fraction' ($f_{\rm{deV}}$). In SDSS, the composite model $r$-band flux for a galaxy, $F_{\rm{comp}}$, is calculated by taking a best-fit linear combination of flux in a de Vaucouleurs profile with that in an exponential profile, such that $F_{\rm{comp}} = f_{\rm{deV}} F_{\rm{deV}} + (1 - f_{\rm{deV}}) F_{\rm{exp}}$, i.e. $F_{\rm{comp}} = F_{\rm{deV}}$ and $f_{\rm{deV}} = 1$ for the most bulge dominated cases.

Whilst defining ellipticals based on this parameter alone is sufficient for most applications, if we are to estimate the typical star formation of ellipticals we must take particular care to avoid any contamination from non-ellipticals, which could have a significant weighting on such calculations.

To estimate the contamination when classifying on $f_{\rm{deV}}$ alone, we cross-match, within 2.5", the galaxy sample with Galaxy Zoo 1 \citep[combined main and bias studies;][]{LIN11}, before isolating those $z < 0.2$, $r < 20$ mag galaxies (although most have $r < 18$ mag) with {$P_{\rm{E,GZ}} > 0.8$} (4692 galaxies) and those $z < 0.2$, $r < 20$ mag galaxies with {$P_{\rm{E,GZ}} < 0.2$} (3874 galaxies), where $P_{\rm{E,GZ}}$ is the fraction of votes for an elliptical classification within the Galaxy Zoo Project. 

Of those confidently classified galaxies with either {$P_{\rm{E,GZ}} > 0.8$} or {$P_{\rm{E,GZ}} < 0.2$}, we find that 20\% of galaxies with $f_{\rm{deV}} > 0.5$ have {$P_{\rm{E,GZ}} < 0.2$}. Imagine this interloping 20\% possess (a maybe even conservative) 10 times the star formation rate of the genuine ellipticals in the sample: Suddenly, $\sim 70$ of all CCSNe (and star formation) within this sample would stem from interlopers.

A naive solution would be to inspect the {$\sim 45000$} $z < 0.2$, $r < 20$ mag, {$f_{\rm{deV}} > 0.5$} galaxies manually to remove non-ellipticals. A consistent inspection of this number of galaxies is not only impractical but we also find that an over-reliance on manual inspection results in a significant redshift bias towards the selection of nearby objects, due to the better resolution of galaxy structure at lower redshifts.

We instead find that a Bayesian classification method, effectively based on distributions of colour and morphology, yields a near-complete sample of ellipticals, whilst removing contaminants effectively, and minimising redshift selection biases.

We treat our previously obtained samples of {$P_{\rm{E,GZ}} > 0.8$} and {$P_{\rm{E,GZ}} < 0.2$} galaxies as respective reference samples of ellipticals and non-ellipticals. We use redshift-debiased values of $P_{\rm{E,GZ}}$ from the Galaxy Zoo project's spectroscopic sample, and the raw classification fractions from their photometric sample. Note, however, that for the low redshift range we are concerned with in the present work, the median change to classification fraction is small, at $\sim 5\%$ for {$0.15 < z < 0.2$}, {$r < 20$ mag} galaxies.

We next arrive at an elliptical classification confidence level, $P_{\rm{E,Bayes}}$, using a range of variables $i,i+1,i+2,...,n$. Let ${P(E\;|\;i)}$ correspond to the probability that a galaxy would belong to the {$P_{\rm{E,GZ}} > 0.8$} sub-sample, given the galaxy's value for a variable $i$. Likewise, let ${P(N\;|\;i)}$ denote the probability that the galaxy would belong to the {$P_{\rm{E,GZ}} < 0.2$} sub-sample given this input. According to Bayesian statistics, the odds ratio ${P(E\;|\;i)}/{P(N\;|\;i)}$ is then given by Equation~\ref{eq:Bayesian1}, where ${P(E)}/{P(N)}$ is the odds ratio prior to the consideration of our variable $i$. Crucially, ${P(i\;|\;E)}/{P(i\;|\;N)}$ equates to the ratio of probability density function heights at the value for $i$.

\begin{equation}\label{eq:Bayesian1}
\frac{P(E\;|\;i)}{P(N\;|\;i)} = \frac{P(i\;|\;E)}{P(i\;|\;N)} \frac{P(E)}{P(N)} 
\end{equation}

We assume as an initial prior that ${P(E)}/{P(N)} = 1$. As the odds ratio becomes the new prior when we combine information from successive variables, it follows that a galaxy's elliptical classification confidence level is given directly from the product of several PDF height ratios, as shown by Equation~\ref{eq:Bayesian2}.

\begin{equation}\label{eq:Bayesian2}
P_{\rm{E,Bayes}} = \left[\left(\prod_{i}^{n} \frac{P(i\;|\;E)}{P(i\;|\;N)}\right)^{-1} + 1\right]^{-1}
\end{equation}

\begin{figure*}
\centerline{\includegraphics[width=1.12\textwidth]{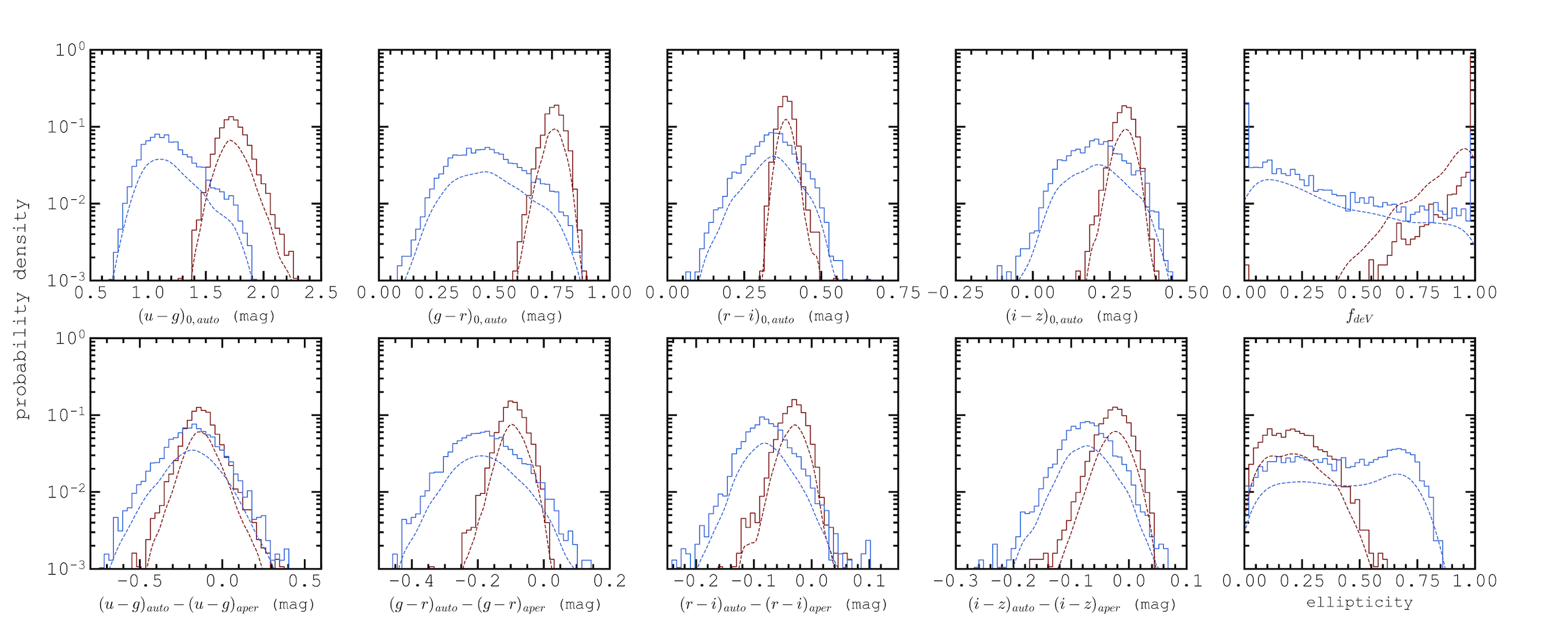}}
\caption{The 10 sets of probability density functions used to classify elliptical galaxies (see text for details). Dark-red series denote the training sample of $P_{\rm{E,GZ}}>0.8$ $z<0.2$ $r<20$ mag galaxies in the Stripe 82 region. Blue series show the same but with instead $P_{\rm{E,GZ}}<0.2$. Gaussian KDEs are fitted to the distributions. For a given galaxy, the ratios of the KDE heights at a given value for each variable are used to calculate $P_{\rm{E,Bayes}}$.}\label{fig:kde} 
\end{figure*}

We use 10 variables in total to calculate $P_{\rm{E,Bayes}}$, as shown in Figure~\ref{fig:kde}. The PDFs used as inputs for Equation~\ref{eq:Bayesian2} equate to Gaussian kernels fitted to each variable's distribution, each smoothed empirically to avoid non-physical discontinuities.

We firstly use rest-frame $ugriz$ colours within elliptical $\textsc{auto}$ apertures, corrected for Galactic extinction using the maps of \citet{SFD98} and $k$-corrected with the prescription of \citet{CHI10}. These primarily help isolate `red sequence' galaxies from the star-forming `blue cloud' \citep[see, e.g.][]{BAL04}. 

We next use the difference between $\textsc{auto}$ aperture colours and those within 2.5" radius circular apertures \citep[denoted \textsc{aper},][]{BA96}. Late-type galaxies typically feature disks which are bluer than their bulges, due to the higher star formation levels in the former regions. Conversely, ellipticals display relatively radially consistent colours. As a result, the wider spread about a null colour difference seen for the non-elliptical reference sample helps us to exclude disk galaxies.

We finally use 2 measures of morphology. The first is the apparent ellipticity, measured in the $r$-band. The second is $f_{\rm{deV}}$. These 2 morphological parameters primarily aid the exclusion of dusty, red, star-formers.

\begin{figure*}
\centerline{\includegraphics[width=0.85\textwidth]{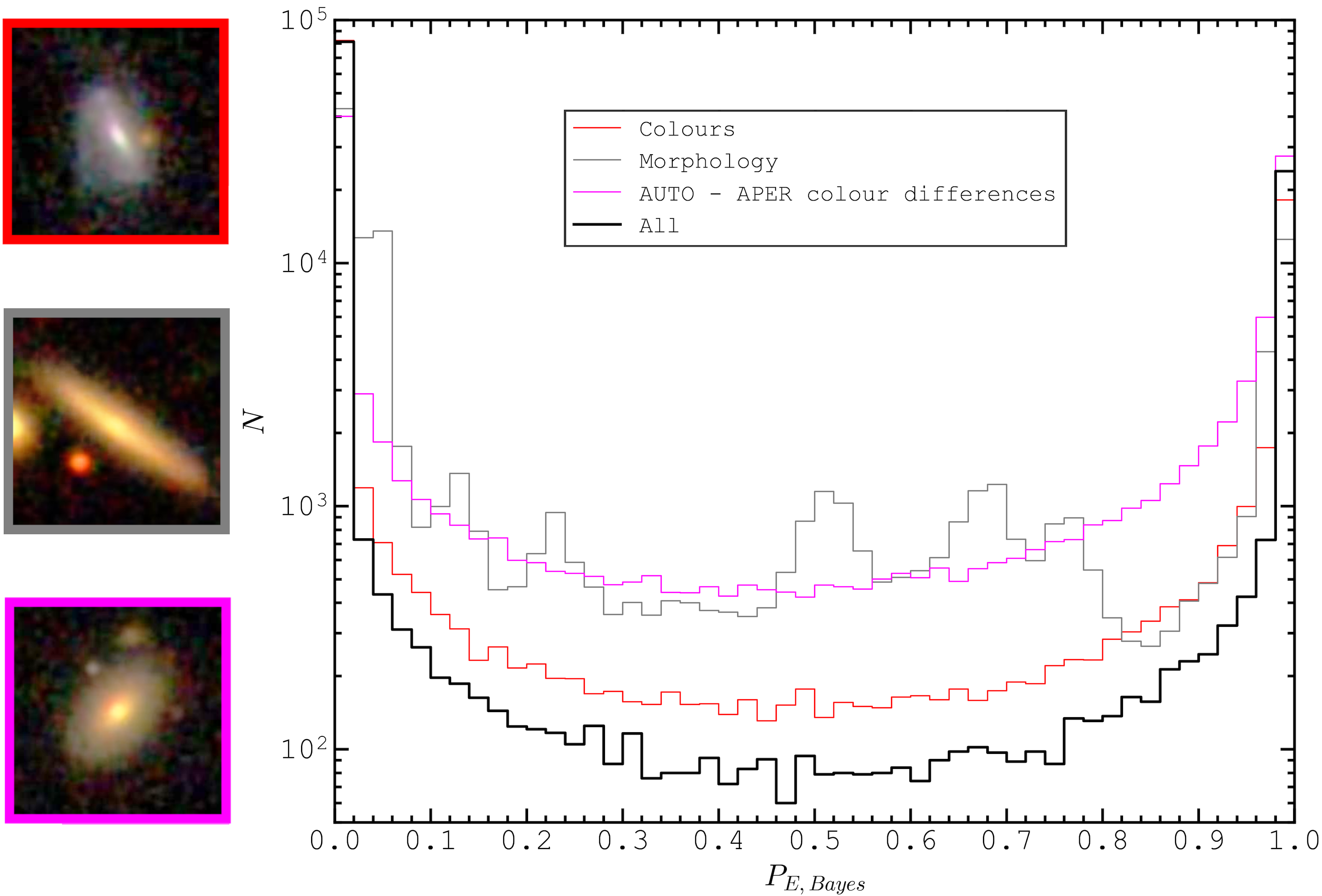}}
\caption{Distributions of $P_{\rm{E,Bayes}}$ values for the sample of $\sim 10^5$ $z<0.2$ $r<20$ mag Stripe 82 galaxies. In black: the final $P_{\rm{E,Bayes}}$ values using all 10 variables in unison. In red, grey, and magenta: $P_{\rm{E,Bayes}}$ values from galaxy colours, morphology (ellipticity and $f_{\rm{deV}}$), and $\textsc{auto}$-$\textsc{aper}$ colour differences alone, respectively. Adjacent images show examples of galaxies with final values of $P_{\rm{E,Bayes}}$ < 0.5, but which would have had $P_{\rm{E,Bayes}}$ > 0.5 were it not for a single set of variables, where image border colour denotes the set of variables which caused the crucial reduction to $P_{\rm{E,Bayes}}$.}\label{fig:P_E_bayes_dists}
\end{figure*}

The distribution of final $P_{\rm{E,Bayes}}$ values found from the 10 sets of input PDFs in unison is shown as the black series in Figure~\ref{fig:P_E_bayes_dists}. Approximately {$ 95\%$} of all $z < 0.2$, $r < 20$ mag galaxies have $P_{\rm{E,Bayes}}$ > 0.95 or $P_{\rm{E,Bayes}}$ < 0.05. This includes all galaxies, not just the training sample.

Also shown are the distributions of $P_{\rm{E,Bayes}}$ values resulting from the parameters used in isolation. It can be seen that galaxy colours are the most crucial parameters for a confident elliptical classification, with $\sim 90 \%$ of $z < 0.2$, $r < 20$ mag galaxies having $P_{\rm{E,Bayes}}$ > 0.95 or $P_{\rm{E,Bayes}}$ < 0.05 from colour alone. Morphology ($f_{\rm{deV}}$ and ellipticity) and $\textsc{auto}$-$\textsc{aper}$ colour differences are of secondary and comparable importance for classification confidence. 

Of course, confidence is not always a reflection of the accuracy of classifications: Each of the parameters helps remove a different sort of contaminating object, and must be used in unison for an effective isolation of ellipticals. This is emphasised with the Stripe 82 coadded images included in Figure~\ref{fig:P_E_bayes_dists}, which show galaxies that would have been assigned {$P_{\rm{E,Bayes}}$ > 0.5} were it not for a given criterion. 

The top image shows a nearby irregular galaxy which exhibits similar morphological properties to a massive elliptical, but was rejected after a consideration of its optical colours. The central image shows an edge-on, dusty, star-forming galaxy which is consistent in colour with an elliptical, but was rejected due to $f_{\rm{deV}}$ and ellipticity. The bottom image shows either a poorly resolved spiral or a lenticular disk, consistent in both colour and morphology with an elliptical, yet rejected due to its $\textsc{auto}$-$\textsc{aper}$ colour differences.

Of a total of 113239 $z < 0.2$, $r < 20$ mag galaxies, 27940 ($\sim 25\%$) have {$P_{\rm{E,Bayes}}$ > 0.5}, and it is these galaxies which define the elliptical sample, to be used in successive sections of the present work.

\begin{figure*}
\centerline{\includegraphics[width=\textwidth]{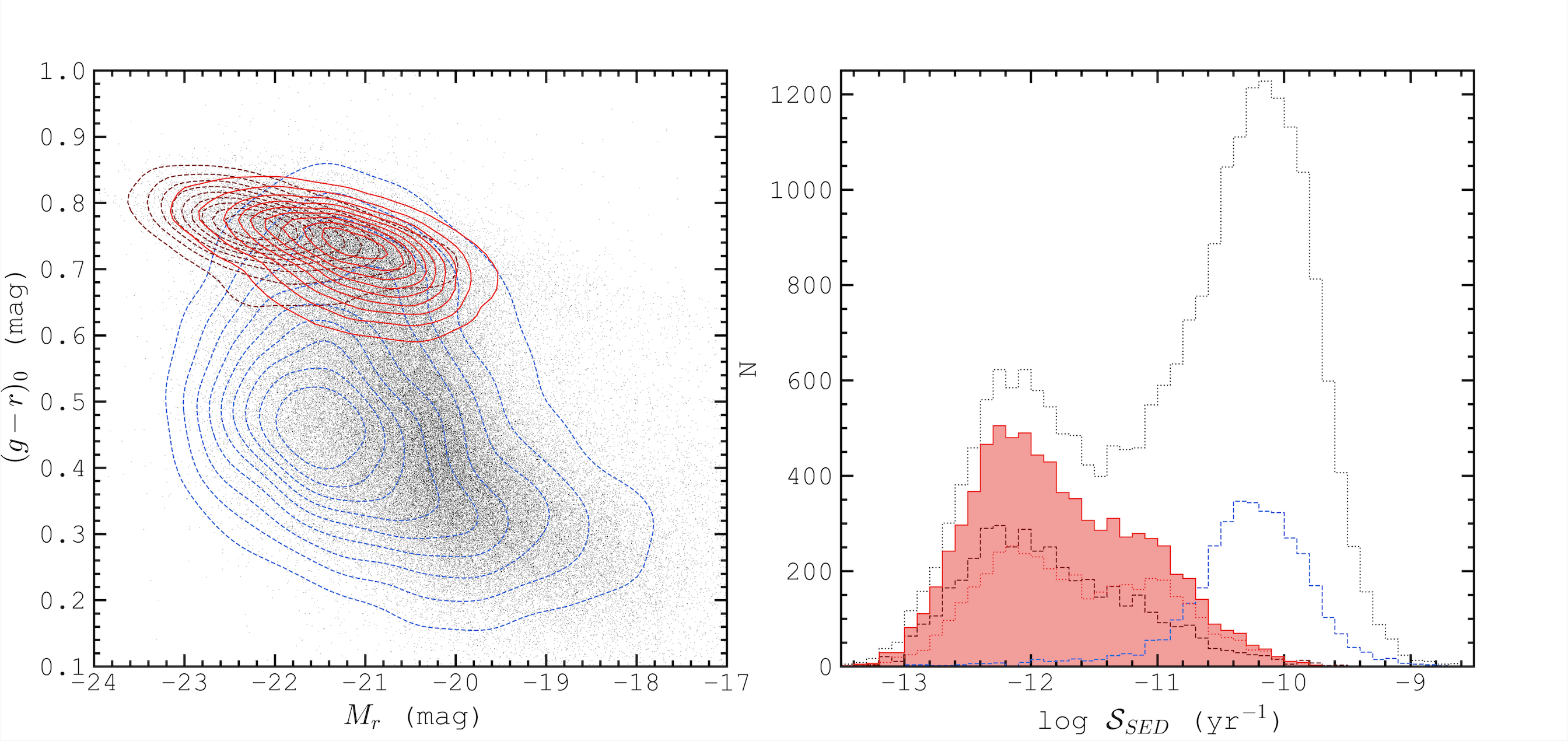}}
\caption{Left: Rest-frame $r$-band absolute magnitude versus rest-frame ($g-r$) colour, where all magnitudes are those within elliptical $\textsc{auto}$ (`Kron') apertures and are corrected for Galactic extinction. Contour levels are 10\% to 90\% of the peak number density in steps of 10\%. Dashed-dark-red and dashed-blue contours depict the training samples of {$P_{\rm{E,GZ}}$ > 0.8} and {$P_{\rm{E,GZ}}$ < 0.2} galaxies, respectively. The solid-red contours represent the sample of ellipticals resulting from our Bayesian classification method. Black points show the distribution of all $z < 0.2$, $r < 20$ mag Stripe 82 galaxies. Right: Galaxy counts as a function of specific-star formation rate as measured by GSWLC-2. The dashed-dark-red, dashed-blue, red-filled and black-dotted series represent the same respective sub-samples, described for the left-hand panel. The red-dotted series depicts the same sample as shown by the red-filled series, but excluding the training set of {$P_{\rm{E,GZ}}$ > 0.8} galaxies.}\label{fig:sample_tests}
\end{figure*}

Manual inspection of these objects reveals little to no contamination from any obvious non-ellipticals, nor any obvious evidence of LSBGs along the line-of-sight to these ellipticals which could have instead housed the observed CCSNe. Figure~\ref{fig:sample_tests} shows 2 more quantitative tests for the validity of our classifications. The left-hand panel shows $k$-corrected $r$-band absolute magnitude versus rest-frame ($g-r$) colour, where all magnitudes are in $\textsc{auto}$ apertures, with the aperture defined in the $r$-band. Contour levels are 10\% to 90\% of the peak number density in steps of 10\%. The fact that the dashed-red and dashed-blue contours overlap shows that, much like a rudimentary cut on $f_{\rm{deV}}$, a `hard' cut on colour would not be able to remove non-ellipticals without also removing a significant number of ellipticals.

We see that our sample of ellipticals resulting from the Bayesian classification method (solid-red) follows a similar distribution to the training set of {$P_{\rm{E,GZ}}$ > 0.8} galaxies (dashed-dark-red), with the only notable difference being an offset in absolute magnitude of $\sim 1$ mag between the peak density of the 2 distributions. This is due to the bias of having more confident manual classifications in Galaxy Zoo for brighter galaxies. This is also clear from the fact that the magnitude at which we find peak number density in the {$P_{\rm{E,GZ}}$ < 0.2} sample (dashed-blue) does not correspond the magnitude at which peak density is seen for all galaxies (black points).

As a comparison with the CCSN-derived star formation results of the present work we will repeatedly compare with star formation rates derived from UV/Optical SED fitting applied to the second edition of the GALEX-SDSS-WISE Legacy Catalogue \citep[GSWLC-2;][]{SAL16,SAL18}. 

In the present work, 23009 out of 113239 $z < 0.2$ $r < 20$ mag galaxies are matched with the GSWLC-2 sample within 2.5". These matches are {$r_{\rm{petro}}$ < 18 mag} galaxies with a spectroscopic redshift in the range {0.03 < $z$ < 0.2}, which lie within the GALEX footprint \citep{MAR05,MOR07}.

Star formation properties of these galaxies were estimated using SED fitting with the Code Investigating GALaxy Emission \citep[\textsc{cigale};][]{NOL09,BOQ19}. Excellent summaries of the SED fitting technique are given in \citet{SAL18} and \citet{TUR21}, but in short: Synthetic spectra, generated using the simple stellar population templates of \citet{BC03}, based on a \citet{CHA03} initial mass function and covering a wide range of metallicities {($\log (Z) = -2.4$ to $-1.3$)}, are fitted to the observed UV-to-Optical photometry.

Templates were then combined with Myr-resolution star formation histories. The library of these histories were built using exponentials with various decay times for an old stellar population, with a relatively flat burst superimposed (at least 100 Myr ago) to represent a younger population. The SED estimates of star formation rate were additionally constrained by the galaxy IR luminosity, itself calculated by matching the energy absorbed by a galaxy's dust with the energy it re-emits. Other properties derived in this fitting procedure include (but are not limited to) galaxy stellar masses, stellar ages, stellar metallicities, absolute magnitudes and colour excess. In the present work, we use the SFRs derived using the deepest photometry available for each galaxy (catalogue GSWLC-X2).

The right-hand panel of Figure~\ref{fig:sample_tests} shows the matched distribution of SED-derived SSFRs ($\mathcal{S}_{\rm{SED}}$). It is seen that the peak value of $\mathcal{S}_{\rm{\rm{SED}}}$ comes at $\sim 10^{-12.0}$ yr$^{-1}$ for both the training sample and Bayesian-classified sample of ellipticals. A notable difference between our final elliptical sample and the Galaxy Zoo training sample is that the former exhibits a secondary peak at $\mathcal{S}_{\rm{SED}}$ $\sim 10^{-11.0}$ yr$^{-1}$. However, we find no significant difference in the appearance of the galaxies classified as ellipticals at these 2 different levels of star formation, and so the origin of this secondary peak remains unexplained.

According to our classifications, 74\% of $\mathcal{S}_{\rm{SED}} < 10^{-11.0}$ yr$^{-1}$ are classified as ellipticals. This number rises to 80\% for $\mathcal{S}_{\rm{SED}} < 10^{-12.0}$ yr$^{-1}$. Ellipticals become the dominant class for $\mathcal{S}_{\rm{SED}} \lesssim 10^{-10.8}$ yr$^{-1}$.

In summary, this Bayesian method is able to isolate a near-complete sample of $z < 0.2$, $r < 20$ mag elliptical galaxies in the Stripe 82 region. In theory, this method could be extended to efficiently classify several species of galaxies over wide-field survey footprints, in a manner consistent over redshift.

\section{The star formation rate density in ellipticals}\label{sec:SFRD}

Our CCSN hosts were classified simultaneously with the larger galaxy sample, and as such are subject to identical classification criteria as in the previous section. Of our sample of 421 likely CCSN hosts ($P_{\rm{Ia}} < 0.5$), 36 are classified as ellipticals using our Bayesian procedure. 27 of these have $P_{\rm{Ia}} < 0.05$.

\begin{figure*}
\centerline{\includegraphics[width=\textwidth]{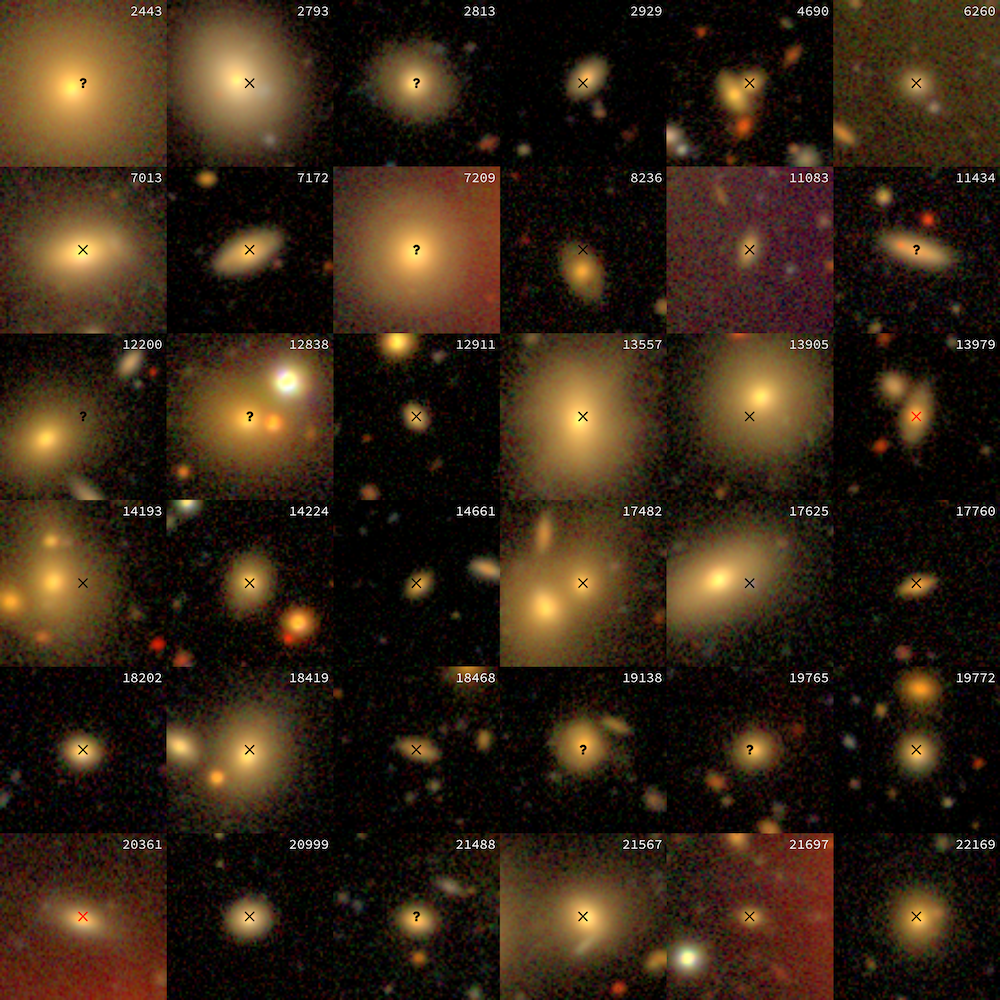}}
\caption{36 $z < 0.2$ elliptical Galaxies found to host $r_{peak} < 21.8$ CCSNe from the SDSS-II Supernovae Survey, as shown in {30" x 30"} coadded images from the IAC Stripe 82 Legacy Survey. Crosses (question marks) denote the locations of CCSNe with $P_{\rm{Ia}} < 0.05$ ($0.05 < P_{\rm{Ia}} < 0.5$). Black (red) symbols denote those with (without) spectra. Numbers indicate SN catalogue IDs from the SDSS-II Supernova Survey.}\label{fig:host_stamps}
\end{figure*}

Co-added Stripe 82 images of these 36 galaxies are shown in Figure~\ref{fig:host_stamps}. These host galaxies can now be used to estimate the contribution to the cosmic star-formation density from ellipticals. 
We can utilise Equation~\ref{eq:ccsnrd} to estimate the CCSN-rate density per unit volume $V$ in ellipticals using the sample of elliptical CCSN host galaxies (see also \citealt{SED19a}). Galaxy stellar masses are first calculated using the same prescription as in \citet{SED19a}; a k-correction inclusive prescription based on $i$-band $\textsc{auto}$ mag, ($g-i$) observed colour (in 2.5" circular apertures) and redshift.

To correct for the fact that our SN sample is flux-limited but we instead want volume-limited CCSN statistics, we implement a statistical correction identical to that used in \citet{SED19a}. This is similar to a 1/$V_{\rm{max}}$ method, but where the SN light is the determining factor in detection, not the host galaxy light. We employ the volume-limited absolute magnitude distributions of \citet{R14} for Type Ib/c and Type II SNe. For each SN, $i$, in our own sample, we use the mean of the absolute magnitude distribution for its SN type $j$, denoted $\overline{M}_{j}$. We can then compute the mean \textit{expected} apparent magnitude for our SN, $\overline{m}_{i}$, as a function of its redshift ($z_i$), Galactic extinction ($A_{r,\rm{MW},i}$), host-galaxy extinction ($A_{r,\rm{h},i}$) and $k$-correction ($k_{r,i}$), as shown by Equation~\ref{eq:SNcorrections1}.

\begin{equation}\label{eq:SNcorrections1}
\overline{m}_{i} = \overline{M}_{j} + 5 \log d_{L}(\mathrm{z_{i}}) + k_{r,i} + A_{r,\rm{h},i} + A_{r,\rm{MW},i}
\end{equation}

Using the standard deviation in the relevant absolute magnitude distribution, $\sigma_{j}$ we can estimate the detectable fraction, $\epsilon_{i}$, of SNe that would have {$r < 21.8$ mag}, for the redshift, extinction and $k$-correction of our SN, using Equation~\ref{eq:SNcorrections2}. The inverse of the fraction $\epsilon_{i}$ is the SN's (and hence its host's) weight of contribution to the volumetric number density.

\begin{equation}\label{eq:SNcorrections2}
\epsilon_{i}=\frac{1}{2} - \frac{1}{2}\erf \left (\frac{\overline{m_{i}}-21.8}{\sqrt{2}\sigma_{j}}\right )
\end{equation}
\begin{equation}\label{eq:ccsnrd}
  \rho_{\rm CCSN}(\mass) \:  = \: \frac{1}{\Delta \log \mass}  \: \frac{\sum_{i}^{} {1}/{\epsilon_{i}(\mass)}}{\tau \, V}
\end{equation}

Equation~\ref{eq:ccsnrd} shows that summing over a mass bin of width $\Delta \log \mass$, the quantity $\sum_{i} {1}/{\epsilon_{i}(\mass)}$ leads to the volume-corrected number of CCSNe associated with galaxies for each bin. $\tau$ is the effective rest-frame time over which CCSNe could be identified by the survey. This is shorter than the observed time-frame of the un-targeted supernova survey, t, such that {$\tau$ = t / (1+$\overline{z}$)}. The observable volume, $V$, is derived from the sky coverage of the Stripe 82 region and the imposed {$z < 0.2$} limit. The relationship between the CCSN rate and SFRD is then given by Equation~\ref{eq:CCSNSFRD}. As derived in \citet{SED19a}, we adopt {$\log \overline{\mathcal{R}} = -1.9$}.

\begin{figure*}
\centerline{\includegraphics[width=0.95\textwidth]{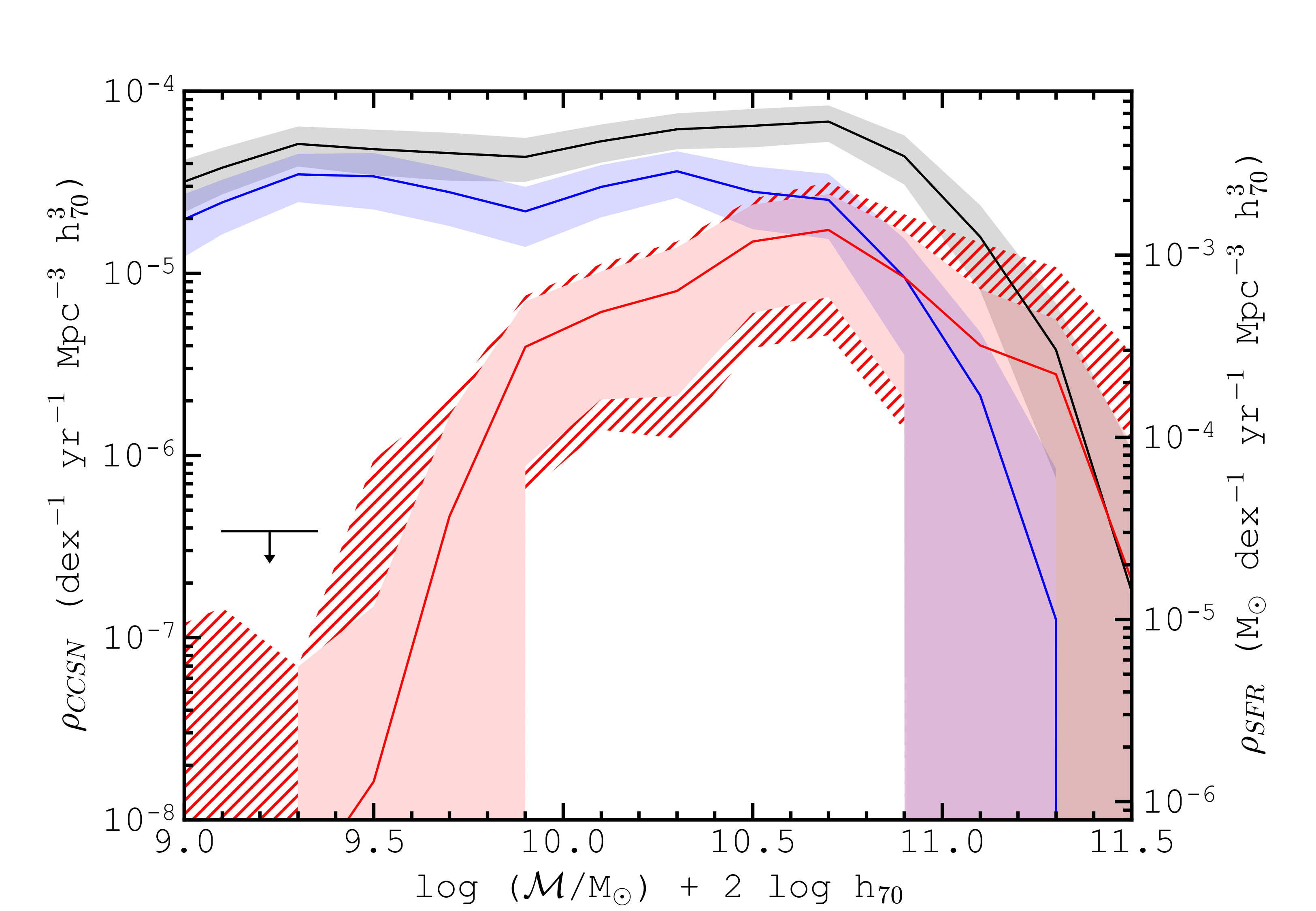}}
\caption{The core-collapse supernova rate density (left-hand vertical axis) and the implied star formation rate density (right-hand vertical axis), as a function of galaxy stellar mass. The red solid line depicts the contribution to these densities from elliptical galaxies. The solid filled red region depicts the associated statistical uncertainty, which is the quadrature sum of MC + Poisson + Cosmic Variance errors. Hatched red regions depict systematic errors, from the treatment of {$0.05 < P_{\rm{Ia}} < 0.5$} objects (positive error) and from the dust extinction assumed for the elliptical hosts (negative error). The blue (black) line and filled region shows the CCSN-rate and SF-rate as a function of mass for blue star-forming disk galaxies (all galaxies). The black arrow shows the upper-limit at {$10^{9.0} - 10^{9.5} \rm{M}_{\odot}$} at which number counts are consistent with zero (see text).}\label{fig:ccsnrd} 
\end{figure*}

We utilise a $10^3$ iteration Monte Carlo (MC) technique to account for uncertainties in redshift, galaxy magnitudes and the resultant stellar masses which all feed into our calculation of the CCSN-rate density and SFRD as a function of galaxy stellar mass, the results for which are shown in Figure~\ref{fig:ccsnrd}. The statistical errors (labelled `stat') on these densities are represented by the filled regions, and equate to the quadrature sum of MC errors (standard deviation of the densities over MC iterations), Poisson errors and the cosmic variance, the latter estimated given the volume of the Stripe 82 region out to $z = 0.2$ \citep{DR10}. The red series denotes the estimate of CCSN-rate and star formation rate density in ellipticals as a function of mass.

The black arrow shows the approximate upper limit to counts for $10^{9.0} - 10^{9.5} \rm{M}_{\odot}$. This is the appropriate one-sided {$1 \sigma$} error on bins with zero counts \citep{GEH86}, normalised for the SN survey coverage, $\tau \times V$, as well as the mean detection efficiency and redshift for the aforementioned mass range. 

Additionally there are 2 main sources of systematic uncertainty on this result. The first relates to uncertainties in SN classifications within the SN Sample: The CCSN-rate density described above is formed from the SN sample excluding those with $P_{\rm{Ia}} > 0.05$. Our positive systematic uncertainty equates to the increase to densities found when including {$0.05 < P_{\rm{Ia}} < 0.5$} CCSNe.

A second systematic relates to the dust extinction assumed within the host galaxies, or more specifically, the dust screen in the region of the SNe. Equations~\ref{eq:SNcorrections1} and \ref{eq:SNcorrections2} rely on an estimate of this extinction level. From the full CCSN sample of \citet{SED19a}, the mean $r$-band extinction in the SN regions was estimated at 0.5 mag. We therefore also assume a mean extinction of 0.5 mag for elliptical hosts in the present work. However, as the sample in \citet{SED19a} consisted mostly of star-forming galaxies, which can be abundant in dusty star-forming regions, the level of dust attenuation in elliptical hosts may be lower. Indeed. the median Galactic-extinction corrected absolute $r$-band magnitude of $P_{\rm{Ia}} < 0.05$ CCSNe at peak light is {$-17.70$ mag} in ellipticals, compared with {$-17.39$ mag} in blue disks. This implies a median $r$-band extinction of {$\sim$ 0.2 mag} in ellipticals. For comparison, we find a difference of {$0.16$ mag} in median $A_{v}$ estimates when moving from blue disk Type Ia SN hosts (median of {$A_{v} = 0.38$} mag) to elliptical Type Ia SN hosts (median of {$A_{v} = 0.22$} mag), where $V$-band extinction values were estimated by \citet{SAK18} from a MLCS2k2 SN light curve fitting technique \citep{JHA07}. Despite these numbers we use a more conservative negative systematic uncertainty on the elliptical SFRD, which represents the decrease to densities assuming \textbf{zero} extinction in the SN regions. Systematic uncertainties, (labelled `sys') are shown as the hatched regions in Figure~\ref{fig:ccsnrd}.

The SFRD in ellipticals is constrained above zero for a mass range 10$^{9.8}$ - 10$^{11.2}$ M$_{\odot}$. The peak contribution most likely comes between 10$^{10.6}$ - 10$^{10.8}$ M$_{\odot}$, where ellipticals contribute {$1.6 \pm 0.8$ (stat) $^{+0.3}_{-0.7}$ (sys) $\times 10^{-5}$ CCSNe yr$^{-1}$ Mpc$^{-3}$ dex$^{-1}$ h$_{70}^{3}$}, corresponding to an SFRD of {$1.3 \pm 0.6$ (stat) $^{+0.2}_{-0.6}$ (sys) $\times 10^{-3}$ M$_{\odot}$ yr$^{-1}$ Mpc$^{-3}$ dex$^{-1}$ h$_{70}^{3}$} at these masses. All the results of the present work are derived using a 737 cosmology {($h=0.7$, $\Omega_{\rm m}=0.3$, $\Omega_{\Lambda}=0.7$)}.

There is a narrow range of masses which contribute non-negligibly to the SFRD in ellipticals, with $\sim$50\% of the contribution coming for 10$^{10.4}$ - 10$^{10.8}$ M$_{\odot}$, and $\sim$90\% of the contribution found between 10$^{10.0}$ - 10$^{11.2}$ M$_{\odot}$. The integrated density above {10$^{10.0}$ M$_{\odot}$}, corresponding to what is effectively the total elliptical population, is found to be $1.3 \pm 0.3$ (stat) $^{+0.4}_{-0.5}$ (sys) $\times 10^{-5}$ CCSNe yr$^{-1}$ Mpc$^{-3}$ h$_{70}^{3}$, equating to $1.0 \pm 0.3$ (stat) $^{+0.3}_{-0.4}$ (stat) $\times 10^{-3}$ M$_{\odot}$ yr$^{-1}$ Mpc$^{-3}$ h$_{70}^{3}$.

For a comparison with the elliptical SFRD we show the result from the sample of CCSNe found in star-forming `blue disk' galaxies (shown in blue in Figure~\ref{fig:ccsnrd} and in successive figures), which we define as those {$z < 0.2$} galaxies which have $(g-r)_{0}$ < 0.65 mag, $f_{\rm{deV}} < 0.5$, and were found to be non-ellipticals using our classification procedure. These galaxies are expected to be the dominant constituents of the star-forming main sequence.

The SFRD in blue disks as integrated above 10$^{10.0}$ M$_{\odot}$ is a factor {$2.1 \pm 0.7$ (stat) $^{+1.3}_{-0.5}$ (sys)} times that in ellipticals. The SFRD of ellipticals most likely surpasses that in blue star-forming disks above 10$^{10.9}$ M$_{\odot}$, due mainly to the domination of the total galaxy number density from ellipticals at the highest galaxy masses.

We also show, in black, the total SFRD as determined from the SDSS-II SN Sample. Note that the red and blue series do not sum to give the total SFRD: There will be additional contributions from lenticular galaxies, irregular galaxies and dusty star-forming galaxies, which fall in to neither the elliptical or blue disk groups. Our results imply that elliptical galaxies contribute {$11.2 \pm 3.1$ (stat) $^{+3.0}_{-4.2}$ (sys) \%} of the total star-formation budget at present epochs. This rises to {$20.2 \pm 6.0$ (stat) $^{+5.7}_{-7.7}$ (sys) \%} for masses above {10$^{10.0}$ M$_{\odot}$}. Both values are consistent with that of \citet{KAV14a}, who find that $14\%$ pertains to early-type galaxies. We conclude that ellipticals contribute a non-negligible fraction of the total star-formation budget to the present-day cosmic volume.

\section{The specific star formation rates of ellipticals}\label{sec:SSFR}

Equation~\ref{eq:Sbar} demonstrates that the mean SSFR of ellipticals can be derived using our result for the SFRD if we also have a calculation of the GSMF of ellipticals. To calculate the required GSMF, we use a $1/V_{\rm{max}}$ method applied to the sample of 27940 Bayesian classified ellipticals with {$z < 0.2$} and {$r < 20$ mag}. The maximum luminosity distance, $D_{L,max}$, within which each galaxy would remain brighter than our flux limit of {$r = 20$ mag} is given by \begin{equation}
     5 \log (D_{L,max}) = 5 \log (D_{L,obs}) + 20 - r_{\textsc{auto}} + dK \;,
\end{equation}\label{eq:Vmax}where $dK$ is the k-correction for our galaxy at the observed redshift, minus that at the maximum redshift, and where the required maximum redshift is iteratively inferred starting from the value of $D_{L,max}$ with $dK=0$. Finally $D_{L,max}$ values are clipped to lie at or below the inferred luminosity distance at $z = 0.2$. $D_{L,max}$ leads to $1/V_{\rm{max}}$, the values for which act as weights on galaxy number densities in {$\log (\mathcal{M}/\rm{M}_{\odot})$} bins of width 0.2 dex, yielding volume-limited results. We calculate the GSMF of blue disks within our redshift and magnitude limits (a total of 66288 blue disk galaxies), and that of all {$z < 0.2$}, {$r < 20$} mag galaxies (a total of 113239 galaxies), using the same approach.

\begin{figure}
\centerline{\includegraphics[width=1.2\columnwidth]{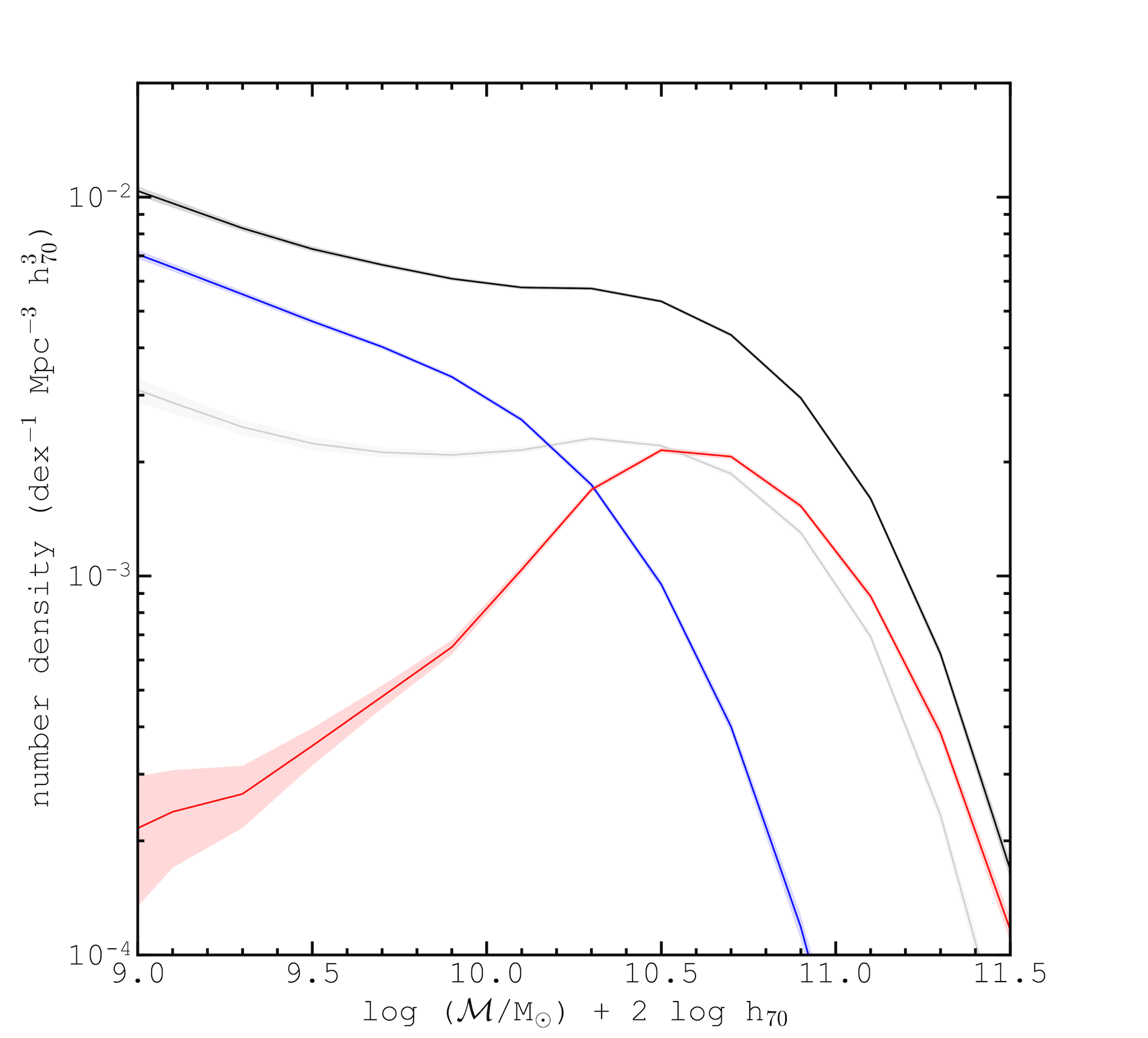}}
\caption{The galaxy stellar mass function. Solid lines and filled regions show $z < 0.2$ galaxy number densities, and their 1~$\sigma$ MC+Poisson+Cosmic Variance errors, respectively, derived from a $1/V_{\rm{max}}$ method using $r < 20$ mag galaxies. In red: elliptical galaxies. In blue: `blue disk' galaxies (see text). In grey: galaxies fitting neither classification. In black: all galaxies.}\label{fig:GSMF}
\end{figure}

Corresponding results are shown in Figure~\ref{fig:GSMF}. Similar to the masses at which the peak of CCSN production is observed, the number density of $z < 0.2$ ellipticals peaks between galaxy stellar masses of $10^{10.4}$ and {$10^{10.8}$ $\rm{M}_{\odot}$}, at {$2.11 \pm 0.03 \times 10^{-3}$ dex$^{-1}$ Mpc$^{-3}$ h$_{70}^{3}$}. Ellipticals dominate the total {$z < 0.2$} galaxy number density for {masses > 10$^{10.8}$ M$_{\odot}$}.

\begin{figure*}
\centerline{\includegraphics[width=1.12\textwidth]{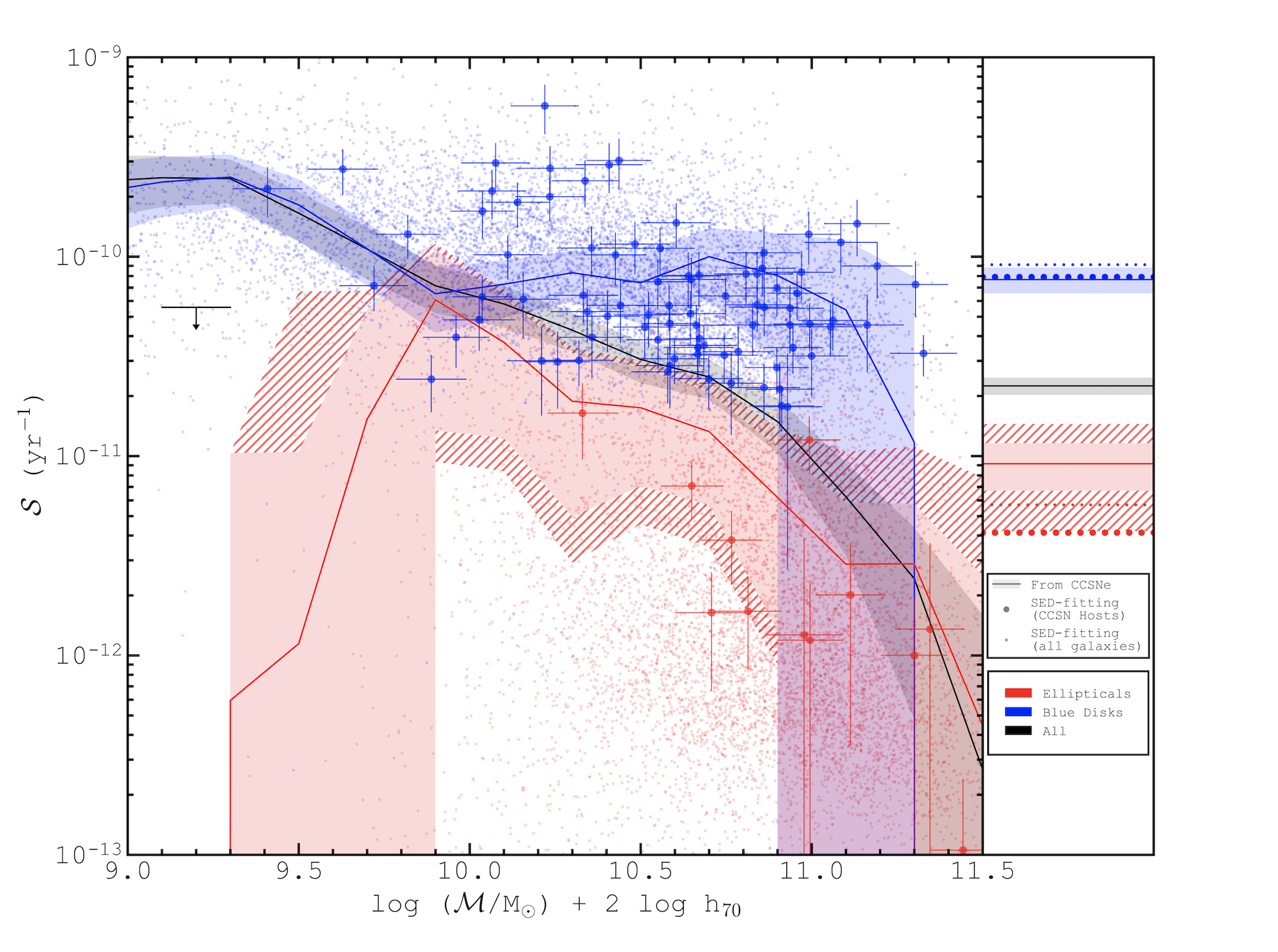}}
\caption{Left: The specific star formation rate as a function of galaxy stellar mass (via multiple means). The red solid line and filled region denote the mean and {1 $\sigma$} uncertainty (MC + Poisson + Cosmic Variance) on the specific star formation rate in elliptical galaxies at $z < 0.2$. Hatched-regions depict additional systematic uncertainties described in the caption of Figure~\ref{fig:ccsnrd}. Small-red points depict the SED-derived specific star formation rates of those \textit{individual} elliptical galaxies cross-matched with GSWLC-X2 data. Large-red points denote cross-matched elliptical CCSN hosts, with error bars showing {1 $\sigma$} errors from SED-fitting. The blue series depict the same but for blue star forming galaxies. The black series shows the mean specific star formation rate as a function of mass for all galaxies. The black arrow shows the upper-limit at {$10^{9.0} - 10^{9.5} \rm{M}_{\odot}$} at which SN counts are consistent with zero (see text). Right: The same results as the left-hand panel but derived using the SFRD and GSMF both integrated over galaxy masses > $10^{10.0}$ M$_{\odot}$.}\label{fig:SSFR}
\end{figure*}

Figure~\ref{fig:SSFR} shows (with the red solid line and filled regions) the mean SSFR of ellipticals as a function of mass, derived from the results for the volumetric SFRD and the GSMF. Using an inverse-square-error weighting on each mass bin, we find that the mean SSFR of ellipticals as a function of mass is well represented by the following regression line for masses {$> 10^{10.0} \rm{M}_{\odot}$}:

\begin{equation}\label{eq:regline_E}
    \log (\overline{\mathcal{S}}(\mathcal{M})/\rm{yr}) = -\; (0.80 \;\pm\;0.59) \; \log (\mathcal{M}/10^{10.5}\rm{M}_{\odot}) \; - 10.83 \;\pm\;0.18 \;.
\end{equation}

The uncertainty on the gradient for this regression line implies there is an $18\%$ chance that $\overline{\mathcal{S}}$ is constant with mass for ellipticals. In relation to this uncertainty, we note that although the estimate of mean SSFR in ellipticals approaches that of main sequence galaxies at {$\log (\mathcal{M}/\rm{M}_{\odot}) \sim 10.0$}, the uncertainty on the result for individual bins is such that $\overline{\mathcal{S}}$ at {$\log (\mathcal{M}/\rm{M}_{\odot}) = 10.0$} could be as low as $10^{-11.0}$ yr$^{-1}$ within $1 \sigma$ errors, in line with the mean for the elliptical population.

Given uncertainties on the SSFR for `blue disks', which effectively define the star-forming main sequence, it is unclear whether the result shows a step function at $\mathcal{M} \sim 10^{9.5} \rm{M}_{\odot}$ and is then flat for higher masses, or whether the SSFR instead exhibits a more gradual negative slope with mass. A slope of some form is likely, given that more massive star-forming galaxies formed earlier and a longer time has passed since the peak of their star formation \citep[see, e.g.][]{GAL05}. 

There may however be reason to expect a step function in SSFR. For instance, \citet{MCG17} find a distinct star forming main sequence for {$8.0 < \log (\mathcal{M}/\rm{M}_{\odot}) < 10.0$} compared to higher masses, with results consistent with constant $\overline{\mathcal{S}}$ in this mass regime. Type II SN-rate models of \citet{GRA15}, which build on the work of \citet{LI11b}, imply a step function in $\overline{\mathcal{S}}$ vs mass centred on {$\log (\mathcal{M}/\rm{M}_{\odot}) \sim 9.5$}. Furthermore, it was found by \citet{SED19a} that assuming a constant SSFR for masses $\lesssim 10^{9.0} \rm{M}_{\odot}$ yields the best consistency between the GSMF from CCSNe and that from a $1/V_{\rm{max}}$ method. Future larger samples of SNe utilised with these SN-based methods will allow us to reach a conclusion for the presence of a step function in $\overline{\mathcal{S}}$ versus mass for main-sequence galaxies \citep[e.g.][]{IVE19}.

The value of $\overline{\mathcal{S}}(\mathcal{M})$ for the total galaxy population could be argued to follow a Schechter parameterisation \citep{SCH76}. Fitting for $\mathcal{M} > 10^{10.0} \rm{M}_{\odot}$, we find best-fit parameters of {[$\log (\mathcal{M}^{*}/\rm{M}_{\odot})$,\;$\overline{\mathcal{S}^{*}}/10^{11.0}$~yr,$\;\alpha$]} = \\{[$10.90 \pm 0.08$,\; $7.94 \pm 2.54$, \;$-1.41 \pm 0.03$].}

Similar to Figure~\ref{fig:ccsnrd}, the black arrow shows the {$1 \sigma$} upper-limit on $\overline{\mathcal{S}}$ for $10^{9.0} - 10^{9.5} \rm{M}_{\odot}$ ellipticals. This value is derived from the one-sided {$1 \sigma$} error on bins shown in Figure~\ref{fig:ccsnrd}, and adopting the average galaxy number density and mass in the aforementioned mass range for the conversion in Equation~\ref{eq:Sbar}. This upper limit shows that, because of the form of Equation~\ref{eq:Sbar}, a fractionally large Poisson error on host galaxy counts leads to a particularly large error on the SSFR at low elliptical masses, spanning several dex. To assess $\overline{\mathcal{S}}$ in ellipticals using CCSNe for $\mathcal{M} < 10^{9.5} \rm{M}_{\odot}$ would require future data from SN surveys of longer time-span and/or larger sky coverage \citep[e.g.][]{IVE19} than those currently available.

Points in Figure~\ref{fig:SSFR} show independent SSFRs for \textit{individual} galaxies derived from UV/Optical SED fitting applied to the GSWLC-2 sample. 7714 of 27940 (8502 of 49641) ellipticals (blue disks) are matched with GSWLC. 12 of 27 (83 of 174) elliptical CCSN hosts (blue disk CCSN hosts) are matched with GSWLC-2, where we here define CCSN hosts as those with $P_{\rm{Ia}} < 0.05$. Hosts are shown as larger points, with their $1\sigma$ errors on $\mathcal{S}_{\rm{SED}}$.

We find that there are no significant differences in the optical colour distributions of the GSWLC-matched samples compared with the full samples of ellipticals and blue disks. We interpret this as a lack of evidence for a bias towards the selection of higher $\mathcal{S}_{\rm{SED}}$ galaxies in the GSWLC-matched sample. As such, a comparison of the CCSN-derived and SED-derived results is constructive.

\begin{figure}
\centerline{\includegraphics[width=1.2\columnwidth]{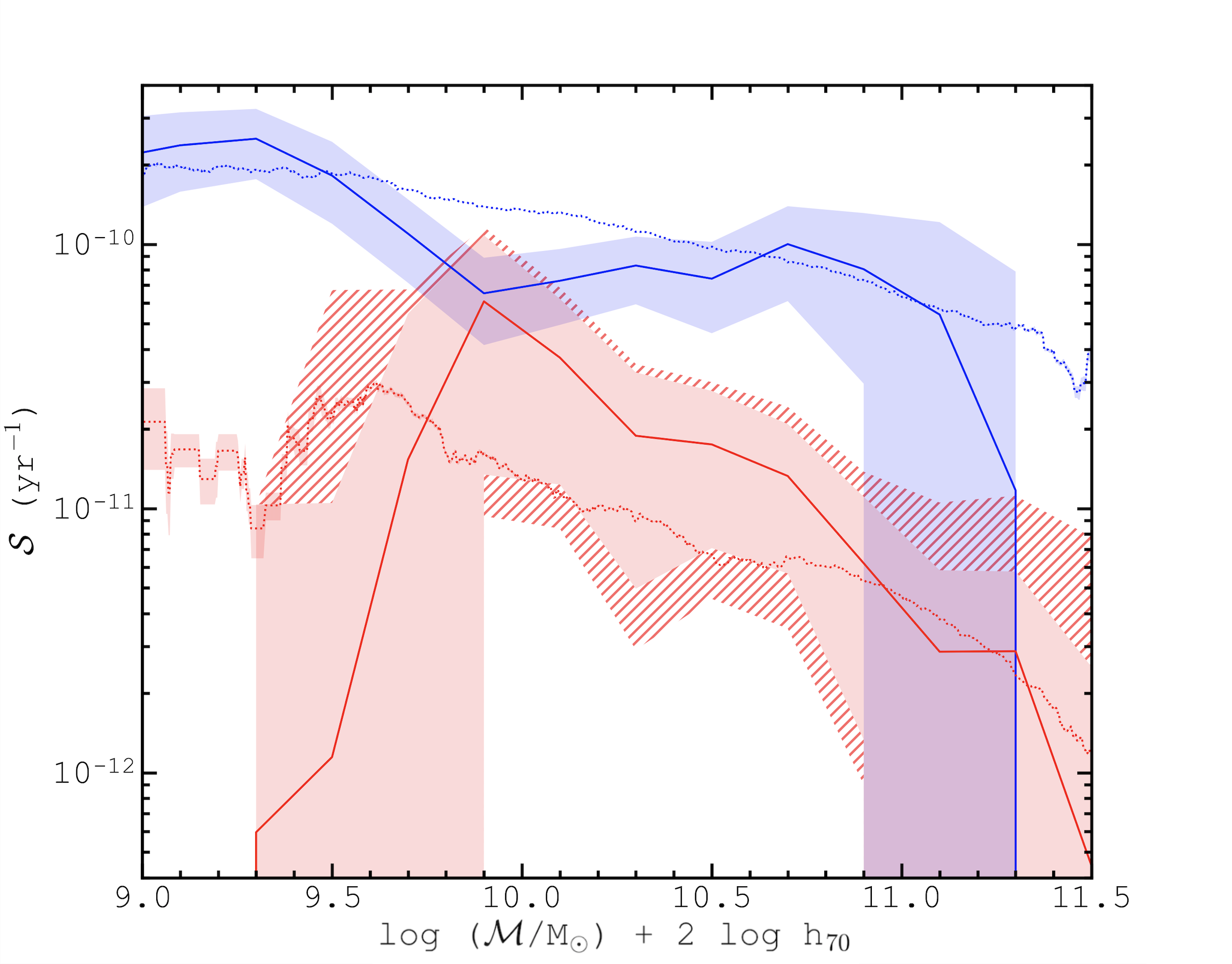}}
\caption{The specific star formation rate as a function of galaxy stellar mass (via multiple means). The red and blue solid lines and surrounding filled regions denote the CCSN-derived results and their uncertainties for elliptical galaxies and blue disk galaxies, respectively, as described in Figure~\ref{fig:SSFR}. The red and blue dotted lines show the running mean of specific star formation rates from SED fitting (derived by \citealt{SAL18}), for elliptical and blue disks, respectively, averaged over 0.1 dex in mass. Surrounding filled regions show the corresponding standard error of these means.}\label{fig:runningmean} 
\end{figure}

$\mathcal{S}_{\rm{SED}}$ values are qualitatively consistent with $\overline{\mathcal{S}}(\mathcal{M})$ derived from CCSNe, with 10 of the 12 GSWLC-matched elliptical CCSN hosts having a measurement of $\mathcal{S}_{\rm{SED}}$ within the $1 \sigma$ uncertainties on the SN-derived $\overline{\mathcal{S}}(\mathcal{M})$, and with the running mean of all GSWLC-matched ellipticals lying within a $1 \sigma$ separation from the SN-derived mean as a function of mass, as shown in Figure~\ref{fig:runningmean}. Interestingly, the running mean value of $\mathcal{S}_{\rm{SED}}$ for the main sequence shows no evidence for the aforementioned step function about $\mathcal{M} = 10^{9.5} \rm{M}_{\odot}$. Furthermore, the gradient of $\mathcal{S}_{\rm{SED}}$ with mass is comparable for both ellipticals and blue disks, which may be in contrast to our findings using CCSNe.

Integrating the volumetric SFRD and the GSMF for {$\mathcal{M} > 10^{10.0} \rm{M}_{\odot}$} (separately) leads to the mean SSFR over this galaxy mass range. Mean values are indicated in the right-hand panel of Figure~\ref{fig:SSFR}. From a CCSN technique, we obtain a value of {$\overline{\mathcal{S}} = 9.2 \pm 2.4$ (stat) $^{+2.7}_{-2.3}$ (sys) $\times 10^{-12}$ yr$^{-1}$} for $\mathcal{M} > 10^{10.0} \rm{M}_{\odot}$ elliptical galaxies.

Also shown is the counterpart value of {$\overline{\mathcal{S}}_{\rm{SED}}$}, taken as the average SED estimate (in linear space) for all GSWLC-matched ellipticals with {$\mathcal{M} > 10^{10.0} \rm{M}_{\odot}$}. We find {$\overline{\mathcal{S}}_{\rm{SED}} = 5.7 \times 10^{-12}$ yr$^{-1}$} for all GSWLC-matched ellipticals, and {$\overline{\mathcal{S}}_{\rm{SED}} = 4.1 \times 10^{-12}$ yr$^{-1}$} for the sub-sample of GSWLC-matched elliptical CCSN hosts. Assuming an average dust extinction in hosts of $A_{r} = 0.5$ mag, these results are in $1.4 \sigma$ tension, and $2.1 \sigma$ tension, respectively, with the value of $\overline{\mathcal{S}}$ measured using a CCSNe. Assuming instead zero dust extinction in elliptical hosts, we find respective discrepancies of $0.5 \sigma$ and $1.1 \sigma$.

The mean SSFR result for {$\mathcal{M} > 10^{10.0} \rm{M}_{\odot}$} blue disks, derived using a CCSN-based method comes to {$\overline{\mathcal{S}} = 7.8 \pm 1.3$ $\times 10^{-11}$ yr$^{-1}$}. This is in $< 1 \sigma$ tension with SED-based measurements, found to be {$\overline{\mathcal{S}}_{\rm{SED}} = 9.1 \times 10^{-11}$ yr$^{-1}$} for all GSWLC-matched blue disks, and {$\overline{\mathcal{S}}_{\rm{SED}} = 7.9 \times 10^{-11}$ yr$^{-1}$} for all matched blue disk CCSN hosts. This consistency reassures our faith in the elliptical result, given that the mean SSFR of the main sequence is well-defined from SED-based measurements (and other methods) within the literature \citep[see, e.g.][]{NOE07,SPE14}. The CCSN-derived measurement of {$\overline{\mathcal{S}}$} for {$\mathcal{M} > 10^{10.0} \rm{M}_{\odot}$} main sequence galaxies implies that ellipticals have a mean SSFR which is {$11.8 \pm 3.7$ (stat) $^{+3.5}_{-2.9}$ (sys) \%} of that on the star-forming main sequence.

\section{The median spectrum of elliptical CCSN hosts}\label{sec:spec}

Note that the SED result averaged over elliptical CCSN hosts is comparable to that averaged over all ellipticals. This implies that whilst CCSNe hosts are indeed probabilistically determined by their star formation rates, the selection function is not so extreme that we are simply tracing the very highest star formation rate ellipticals. This increases confidence that the measurement of $\overline{\mathcal{S}}$ via CCSNe is well-representative of the mean star formation level over the total elliptical population.

Of course, SED-based measurements of star formation rates for ellipticals are subject to uncertainty; which is indeed a major motivation for the present work. We therefore turn to an analysis of the 25 elliptical CCSN host galaxy's coadded spectra, to test whether these hosts exhibit spectral properties typical of the average elliptical. 

Although SED measurements are derived in part from this spectral information, we can observe a more detailed picture of the star formation properties from the direct spectra, and can look at information free of fitting dependencies. 

We first compute the median spectrum of the {$P_{\rm{Ia}} < 0.05$} elliptical CCSN host galaxies for which spectra are available, i.e. those 25 galaxies labelled with black crosses in Figure~\ref{fig:host_stamps}. To do this, each galaxy's rest-frame spectrum is normalised to equal unity when integrated over wavelengths of $4000 \AA$ and $8000 \AA$, before the sample's median normalised flux is taken as a function of wavelength axis at intervals of $0.1 \AA$. The {1~$\sigma$} error on the median flux is calculated from the 16$^{\rm{th}}$ and 84$^{\rm{th}}$ percentiles as a function of wavelength. This uncertainty is shown in each panel except the top main panel, where it is omitted for clarity.

We similarly compute the median spectrum of a control sample of non-hosting ellipticals matched in galaxy stellar mass and redshift to our CCSN hosts. For each host, we find the separation with non-hosts in a Cartesian mass-redshift space, where separations along each axis are normalised by the median separation of random pairs, such that mass and redshift carry comparable weight in the matching procedure. We find the nearest 5 matches for each host elliptical, yielding a control sample of 125 galaxies.

\begin{figure*}
\centerline{\includegraphics[width=1.05\textwidth]{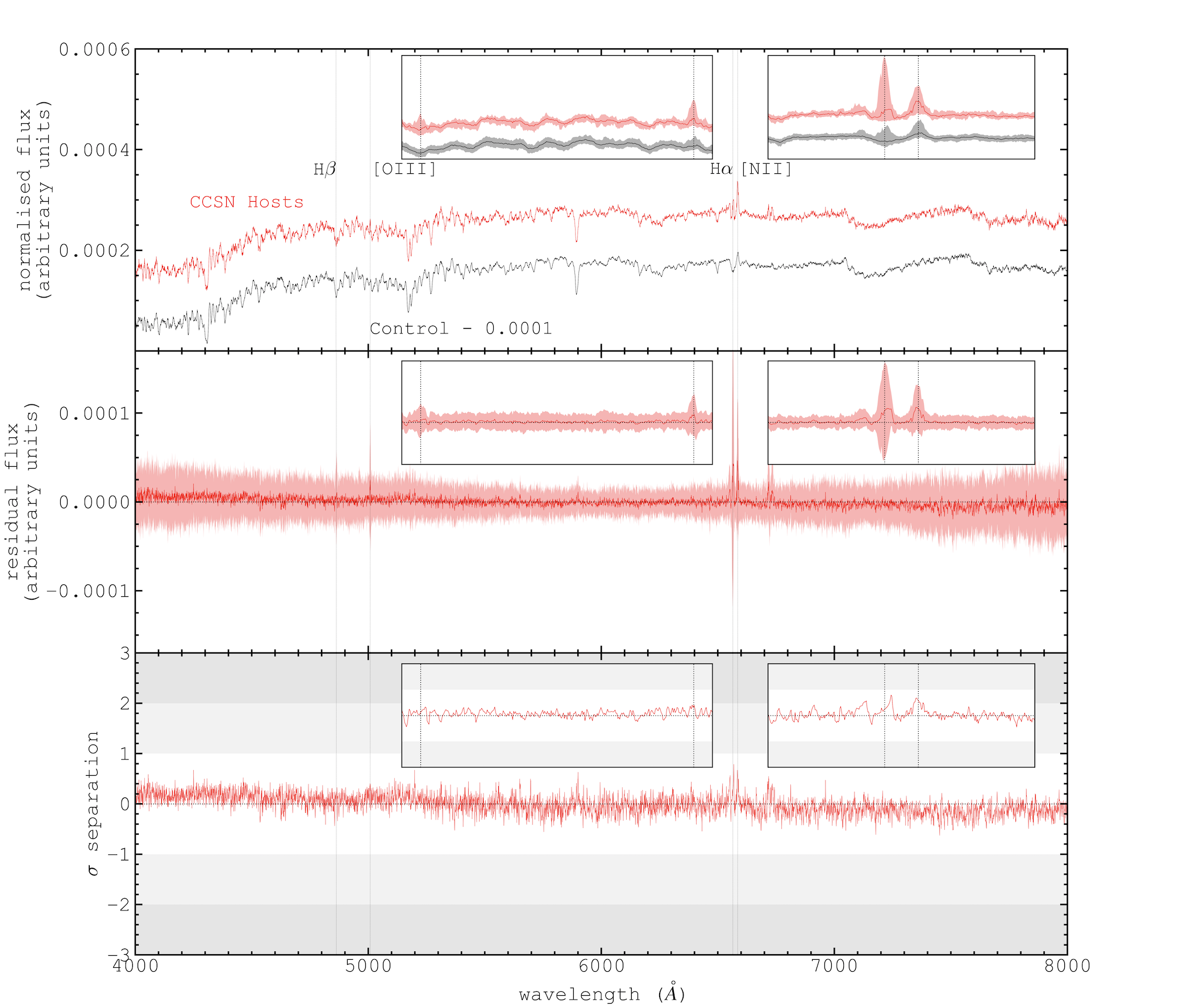}}
\caption{Top panel: The median spectrum of elliptical CCSN hosts (red) vs that of a control sample of (non-hosting) elliptical galaxies (see text). Inset plots show zoomed in wavelength ranges about the H${\beta}$ and H${\alpha}$ lines. Upper and lower bounds on the median normalised flux are signified by filled regions and correspond to the 84$^{\rm{th}}$ and 16$^{\rm{th}}$ percentiles, respectively. These are omitted in the top main panel, for clarity. Middle panel: The residual spectrum (host - control). Bottom panel: The statistical significance of the difference between the median spectra as a function of wavelength.}\label{fig:spectra}
\end{figure*}

As the Balmer lines are notable signatures star-formation properties, we give particular focus to the strengths of H${\alpha}$ and H${\beta}$. We also investigate the ratios of these lines with the respectively nearby [NII]$\lambda6583$ and [OIII]$\lambda5007$ lines, in order to disentangle flux contributions to the Balmer lines from star formation with those from nuclear activity \citep{KAU03}.

As shown in the bottom panel of Figure~\ref{fig:spectra}, we find no statistically significant difference in H${\alpha}$ or H${\beta}$ line strength between the SN host sample and control sample. We also find no statistical difference between the median strengths of [OIII]$\lambda5007$ or [NII]$\lambda6583$. In fact, not one optical emission line is found to differ in its median strength over the 2 samples, within the errors shown.

We also see no difference in the median line ratios [OIII]$\lambda5007$/H${\beta}$ or [NII]$\lambda6583$/H${\alpha}$, implying a similar level of both star formation and nuclear activity for the host sample and control sample. The equivalent width of these 4 emission lines is also consistent across the 2 samples. 

We finally use these line ratios to consider the positions of the 25 elliptical CCSN hosts on the BPT diagram \citep{BPT81}, and find all 25 of these galaxies lie comfortably on the AGN/Passive side of the \citet[][]{KAU03} demarcation. This implies the line ratios are incompatible with high-levels of star-formation and instead the majority contribution to the aforementioned lines is coming from old stellar populations and/or AGN (which would be Type-II AGN in the majority of cases, given the lack of broad lines seen in Figure~\ref{fig:spectra}).

Each of these results implies once more that the elliptical CCSN hosts have typical levels of star formation when compared to the average (or classical interpretation of an) elliptical, and as such we have derived an accurate measure of the mean SSFR of all elliptical galaxies, directly from CCSNe. The consistency between the spectral properties of the CCSN hosts with the remaining elliptical population may also imply that the star formation in ellipticals is not particularly bursty in nature.

\section{Summary}\label{sec:summary}

We have classified all $z < 0.2$, $r < 20$ mag Stripe 82 galaxies using a Bayesian method, trained on the colour and morphology properties of manually-classified samples of ellipticals and non-ellipticals from Galaxy Zoo 1 (Figure~\ref{fig:kde}). Such a concept could certainly be extended to classify other galaxy types, including spiral, lenticular, and irregular galaxies, efficiently over large surveys. In doing so, the connections between Hubble Type and galaxy evolution could be probed with never-before-seen statistical vigour.

We have isolated a sample of 421 $r < 21.8$ mag, $z < 0.2$ likely CCSNe, originating from the SDSS-II Supernova Survey. From their light curves, these SNe each had a Bayesian confidence ($P_{\rm{CC}}$) $> 50\%$ of being a CCSN. 360 ($\sim 86\%$) of these had a Bayesian confidence $> 95\%$, making this a reliable sample of CCSNe. $24 \pm 1 \%$ of these CCSNe are likely to be of Type Ib/c, in line with expectations for the ratio of those of Type Ib/c to Type II. As it is far more common to misclassify Type Ia SNe as Type Ic SNe (lack of light curve plateau and similar spectral properties) rather than as Type II SNe, this reasonable ratio of Type Ib/c SNe to Type II SNe increases our faith in the SN classification procedure. In \citet{SED19a}, these SNe were carefully matched to their host galaxies, meaning we could here isolate those CCSNe residing in ellipticals. 36 CCSNe (27 with $P_{\rm{CC}} > 0.95$) were found to occur in such galaxies (Figure~\ref{fig:host_stamps}).

The $z < 0.2$ volumetric star formation rate density (SFRD; Figure~\ref{fig:ccsnrd}) in elliptical galaxies was calculated from the CCSN rate as a function of host galaxy stellar mass, assuming a ratio between core-collapse events and star formation of {$\overline{\mathcal{R}} = 10^{-1.9}$ $\rm{M}_{\odot}^{-1}$}. The elliptical SFRD result was constrained above zero for ${9.8} < \log (\mathcal{M}/\rm{M}_{\odot}) < 11.2$, with the peak star formation contribution from ellipticals coming at ${10.4} < \log (\mathcal{M}/\rm{M}_{\odot}) < 10.8$. Ellipticals were found to dominate the present-day star formation budget for {$\log (\mathcal{M}/\rm{M}_{\odot}) > 10.9$}.

Ellipticals were found to contribute {$11.2 \pm 3.1$ (stat) $^{+3.0}_{-4.2}$ (sys) \%} of the total star-formation budget. This rises to {$20.2 \pm 6.0$ (stat) $^{+5.7}_{-7.7}$ (sys) \%} for {$\log (\mathcal{M}/\rm{M}_{\odot}) > 10.0$}. These results are consistent with recent studies of star formation in ellipticals using galaxy emission, such as the UV-optical study of \citep{KAV14a}. As CCSN statistics cannot be misconstrued as anything but signatures of recent star formation, we have definitively demonstrated that \textit{ellipticals contribute a non-negligible level of star formation to the present-day cosmic budget}.

The mean specific star formation rate (SSFR; Figure~\ref{fig:SSFR}) of ellipticals was calculated as a function of mass, via the combined information of the previous SFRD result and the galaxy stellar mass function (Figure~\ref{fig:GSMF}). The mean elliptical SSFR most likely exhibits a slope, with a regression line well-representative for galaxy masses $> 10^{10.0} \rm{M}_{\odot}$, such that $\log (\overline{\mathcal{S}}(\mathcal{M})/\rm{yr}) = -\; (0.80 \;\pm\;0.59) \; \log (\mathcal{M}/10^{10.5}\rm{M}_{\odot}) \; - 10.83 \;\pm\;0.18 \;$. However, these errors show there is still an 18\% chance the SSFR is constant with mass in ellipticals. The mean SSFR for all {$\log (\mathcal{M}/\rm{M}_{\odot}) > 10.0$} ellipticals was found to be {$\overline{\mathcal{S}} = 9.2 \pm 2.4$ (stat) $^{+2.7}_{-2.3}$ (sys) $\times 10^{-12}$ yr$^{-1}$}. This is {$11.8 \pm 3.7$ (stat) $^{+3.5}_{-2.9}$ (sys) \%} of the average SSFR on the star-forming main sequence.

An independent mean SSFR in ellipticals derived from GSWLC-2 SED fitting is found to be {$\overline{\mathcal{S}}_{\rm{SED}} = 5.7 \times 10^{-12}$ yr$^{-1}$}, which is moderately consistent with our SN-derived result ($1.4 \sigma$ tension). Assuming zero dust extinction in elliptical CCSN hosts this tension drops to only $0.5 \sigma$. This indicates the SED-derived results put forward as upper limits by \citep{SAL18} are likely close to the true levels of star formation for these objects (See Figure~\ref{fig:runningmean}).

We finally computed the median optical spectrum of the 27 $P_{\rm{Ia}} < 0.05$ CCSN hosts and compared this with that of a control sample of non-hosting ellipticals matched in galaxy stellar mass and redshift (Figure~\ref{fig:spectra}). We find no statistically significant difference in the median strengths of emission lines for these 2 samples, including various commonly associated with star formation (H${\alpha}$, H${\beta}$, [OIII]$\lambda5007$ and [NII]$\lambda6583$). This result implies elliptical CCSN hosts have typical levels of star formation compared to the average elliptical. This in turn signifies that our CCSN-derived results are an accurate representation for the total elliptical population at current epochs. These results are consistent with the hierarchical evolution widely accepted under a $\Lambda$-CDM paradigm, in which low-level star formation is expected to continue in ellipticals out to current epochs, likely due to the influence of galaxy mergers and/or a slow-rate of gas cooling.

Note that out of all red sequence galaxies, CCSNe were only likely to occur in massive red ellipticals within the time-frame of the SDSS-II Supernova Survey, and so cannot reveal information on the star formation properties of lower mass, environmentally quenched galaxies. These may indeed have zero levels of star formation. A relevant investigation may become possible with access to future high-cadence surveys with increased time-span, area and magnitude depth, such as LSST. These surveys would greatly increase the size of CCSN-selected galaxy samples, and would also allow for more precise measurements of the SFRD and SSFR in massive ellipticals. A study such as this with LSST may offer up enough CCSN statistics to allow volume-averaged star formation properties to be estimated as a function of redshift, which would shed light on the average star formation history of ellipticals.

\section{Acknowledgements}

TMS acknowledges support from an STFC DTP studentship, jointly supported by the Faculty of Engineering and Technology at LJMU. SK acknowledges support from the STFC [ST/S00615X/1] and a Senior Research Fellowship from Worcester College Oxford. 

This publication uses data generated via the Zooniverse.org platform, development of which is funded by generous support, including a Global Impact Award from Google, and by a grant from the Alfred P. Sloan Foundation. The authors acknowledge the role of citizen science in the production of Galaxy Zoo data.

The construction of GSWLC was funded through NASA award NNX12AE06G.

Funding for the SDSS-II, SDSS-III and SDSS-IV has been provided by the Alfred P. Sloan Foundation, the National Science Foundation, the U.S. Department of Energy Office of Science, the National Aeronautics and Space Administration, the Japanese Monbukagakusho, the Max Planck Society, the Higher Education Funding Council for England, and the Participating Institutions. SDSS-IV acknowledges support and resources from the Center for High-Performance Computing at the University of Utah. The SDSS web site is www.sdss.org.

SDSS-II, SDSS-III and SDSS-IV are managed by the Astrophysical Research Consortium for the 
Participating Institutions of the SDSS Collaboration including 
the American Museum of Natural History,
University of Arizona, 
Astrophysical Institute Potsdam, 
University of Basel, 
the Brazilian Participation Group, 
Brookhaven National Laboratory, 
University of Cambridge, 
the Carnegie Institution for Science, 
Carnegie Mellon University, 
Case Western Reserve University, 
University of Chicago, 
the Chilean Participation Group,
University of Colorado Boulder,
Drexel University, 
Fermilab, 
University of Florida, 
the French Participation Group,
the German Participation Group,
Harvard-Smithsonian Center for Astrophysics,
Harvard University, 
the Institute for Advanced Study, 
Instituto de Astrof\'isica de Canarias, 
the Japan Participation Group, 
Johns Hopkins University, 
the Joint Institute for Nuclear Astrophysics, 
the Kavli Institute for Particle Astrophysics and Cosmology, 
Kavli Institute for the Physics and Mathematics of the Universe (IPMU) / University of Tokyo, 
the Korean Scientist Group, 
the Chinese Academy of Sciences (LAMOST), 
Lawrence Berkeley National Laboratory, 
Leibniz Institut f\"ur Astrophysik Potsdam (AIP),  
Los Alamos National Laboratory, 
Max-Planck-Institut f\"ur Astronomie (MPIA Heidelberg), 
Max-Planck-Institut f\"ur Astrophysik (MPA Garching), 
Max-Planck-Institut f\"ur Extraterrestrische Physik (MPE), 
the Michigan State/Notre Dame/JINA Participation Group, 
Universidad Nacional Aut\'onoma de M\'exico, 
National Astronomical Observatories of China,
New Mexico State University, 
New York University, 
University of Notre Dame, 
Observat\'ario Nacional / MCTI,
Ohio State University, 
University of Oxford, 
Pennsylvania State University, 
University of Pittsburgh, 
University of Portsmouth, 
Princeton University,
Shanghai Astronomical Observatory, 
the Spanish Participation Group, 
University of Tokyo, 
the United States Naval Observatory, 
University of Utah, 
Vanderbilt University, 
United Kingdom Participation Group,
University of Virginia, 
the University of Washington,
University of Wisconsin, 
Vanderbilt University, 
and Yale University.

\section{Data Availability}

The data underlying this article were accessed from: the Sloan
Digital Sky Survey at {\href{https://www.skyserver.sdss.org}{\url{skyserver.sdss.org}}} (dr14.PhotoPrimary,
dr14.SpecObj, dr14.Photoz, dr7.PhotoObjAll) and at {\href{https://data.sdss.org/sas/dr10/boss/papers/supernova}{\url{data.sdss.org/sas/dr10/boss/papers/supernova}}}; the GALEX-SDSS-WISE Legacy Catalog at {\href{https://salims.pages.iu.edu/gswlc}{\url{salims.pages.iu.edu/gswlc}}}; and the Galaxy Zoo 1 data release at {\href{https://data.galaxyzoo.org}{\url{data.galaxyzoo.org}}}
The data derived in this research, including images, identifiers and properties of the elliptical CCSN hosts, will be shared at {\href{https://www.astro.ljmu.ac.uk/\%7Eikb/research/}{\url{www.astro.ljmu.ac.uk/~ikb/research}}} or on reasonable request to TMS or IKB.
\bibliography{mybib.bib}

\begin{thebibliography}{}
\makeatletter
\relax
\def\mn@urlcharsother{\let\do\@makeother \do\$\do\&\do\#\do\^\do\_\do\%\do\~}
\def\mn@doi{\begingroup\mn@urlcharsother \@ifnextchar [ {\mn@doi@}
  {\mn@doi@[]}}
\def\mn@doi@[#1]#2{\def\@tempa{#1}\ifx\@tempa\@empty \href
  {http://dx.doi.org/#2} {doi:#2}\else \href {http://dx.doi.org/#2} {#1}\fi
  \endgroup}
\def\mn@eprint#1#2{\mn@eprint@#1:#2::\@nil}
\def\mn@eprint@arXiv#1{\href {http://arxiv.org/abs/#1} {{\tt arXiv:#1}}}
\def\mn@eprint@dblp#1{\href {http://dblp.uni-trier.de/rec/bibtex/#1.xml}
  {dblp:#1}}
\def\mn@eprint@#1:#2:#3:#4\@nil{\def\@tempa {#1}\def\@tempb {#2}\def\@tempc
  {#3}\ifx \@tempc \@empty \let \@tempc \@tempb \let \@tempb \@tempa \fi \ifx
  \@tempb \@empty \def\@tempb {arXiv}\fi \@ifundefined
  {mn@eprint@\@tempb}{\@tempb:\@tempc}{\expandafter \expandafter \csname
  mn@eprint@\@tempb\endcsname \expandafter{\@tempc}}}

\bibitem[\protect\citeauthoryear{{Baldry}, {Glazebrook}, {Brinkmann},
  {Ivezi{\'c}}, {Lupton}, {Nichol}  \& {Szalay}}{{Baldry} et~al.}{2004}]{BAL04}
{Baldry} I.~K.,  {Glazebrook} K.,  {Brinkmann} J.,  {Ivezi{\'c}} {\v{Z}}.,
  {Lupton} R.~H.,  {Nichol} R.~C.,   {Szalay} A.~S.,  2004, \mn@doi [\apj]
  {10.1086/380092}, \href
  {https://ui.adsabs.harvard.edu/abs/2004ApJ...600..681B} {600, 681}

\bibitem[\protect\citeauthoryear{{Baldry} et~al.,}{{Baldry}
  et~al.}{2005}]{BAL05}
{Baldry} I.~K.,  et~al., 2005, \mn@doi [\mnras]
  {10.1111/j.1365-2966.2005.08799.x}, \href
  {https://ui.adsabs.harvard.edu/abs/2005MNRAS.358..441B} {358, 441}

\bibitem[\protect\citeauthoryear{{Baldry}, {Sullivan}, {Rani}  \&
  {Turner}}{{Baldry} et~al.}{2021}]{BAL21}
{Baldry} I.~K.,  {Sullivan} T.,  {Rani} R.,   {Turner} S.,  2021, \mn@doi
  [\mnras] {10.1093/mnras/staa3327}, \href
  {https://ui.adsabs.harvard.edu/abs/2021MNRAS.500.1557B} {500, 1557}

\bibitem[\protect\citeauthoryear{{Baldwin}, {Phillips}  \&
  {Terlevich}}{{Baldwin} et~al.}{1981}]{BPT81}
{Baldwin} J.~A.,  {Phillips} M.~M.,   {Terlevich} R.,  1981, \mn@doi [\pasp]
  {10.1086/130766}, \href
  {https://ui.adsabs.harvard.edu/abs/1981PASP...93....5B} {93, 5}

\bibitem[\protect\citeauthoryear{{Bertin} \& {Arnouts}}{{Bertin} \&
  {Arnouts}}{1996}]{BA96}
{Bertin} E.,  {Arnouts} S.,  1996, \mn@doi [\aaps] {10.1051/aas:1996164}, \href
  {https://ui.adsabs.harvard.edu/abs/1996A&AS..117..393B} {117, 393}

\bibitem[\protect\citeauthoryear{{Boquien}, {Burgarella}, {Roehlly}, {Buat},
  {Ciesla}, {Corre}, {Inoue}  \& {Salas}}{{Boquien} et~al.}{2019}]{BOQ19}
{Boquien} M.,  {Burgarella} D.,  {Roehlly} Y.,  {Buat} V.,  {Ciesla} L.,
  {Corre} D.,  {Inoue} A.~K.,   {Salas} H.,  2019, \mn@doi [\aap]
  {10.1051/0004-6361/201834156}, \href
  {https://ui.adsabs.harvard.edu/abs/2019A&A...622A.103B} {622, A103}

\bibitem[\protect\citeauthoryear{{Botticella} et~al.,}{{Botticella}
  et~al.}{2017}]{BOT17}
{Botticella} M.~T.,  et~al., 2017, \mn@doi [\aap]
  {10.1051/0004-6361/201629432}, \href
  {https://ui.adsabs.harvard.edu/abs/2017A&A...598A..50B} {598, A50}

\bibitem[\protect\citeauthoryear{{Bouwens} et~al.,}{{Bouwens}
  et~al.}{2009}]{BOU09}
{Bouwens} R.~J.,  et~al., 2009, \mn@doi [\apj] {10.1088/0004-637X/705/1/936},
  \href {https://ui.adsabs.harvard.edu/abs/2009ApJ...705..936B} {705, 936}

\bibitem[\protect\citeauthoryear{{Bower}, {Lucey}  \& {Ellis}}{{Bower}
  et~al.}{1992a}]{BLE92}
{Bower} R.~G.,  {Lucey} J.~R.,   {Ellis} R.~S.,  1992a, \mn@doi [\mnras]
  {10.1093/mnras/254.4.589}, \href
  {https://ui.adsabs.harvard.edu/abs/1992MNRAS.254..589B} {254, 589}

\bibitem[\protect\citeauthoryear{{Bower}, {Lucey}  \& {Ellis}}{{Bower}
  et~al.}{1992b}]{BOW92}
{Bower} R.~G.,  {Lucey} J.~R.,   {Ellis} R.~S.,  1992b, \mn@doi [\mnras]
  {10.1093/mnras/254.4.601}, \href
  {https://ui.adsabs.harvard.edu/abs/1992MNRAS.254..601B} {254, 601}

\bibitem[\protect\citeauthoryear{{Bruzual} \& {Charlot}}{{Bruzual} \&
  {Charlot}}{2003}]{BC03}
{Bruzual} G.,  {Charlot} S.,  2003, \mn@doi [\mnras]
  {10.1046/j.1365-8711.2003.06897.x}, \href
  {https://ui.adsabs.harvard.edu/abs/2003MNRAS.344.1000B} {344, 1000}

\bibitem[\protect\citeauthoryear{{Butsky}, {Fielding}, {Hayward}, {Hummels},
  {Quinn}  \& {Werk}}{{Butsky} et~al.}{2020}]{Butsky2020}
{Butsky} I.~S.,  {Fielding} D.~B.,  {Hayward} C.~C.,  {Hummels} C.~B.,  {Quinn}
  T.~R.,   {Werk} J.~K.,  2020, \mn@doi [\apj] {10.3847/1538-4357/abbad2},
  \href {https://ui.adsabs.harvard.edu/abs/2020ApJ...903...77B} {903, 77}

\bibitem[\protect\citeauthoryear{{Calzetti}, {Armus}, {Bohlin}, {Kinney},
  {Koornneef}  \& {Storchi-Bergmann}}{{Calzetti} et~al.}{2000}]{CAL00}
{Calzetti} D.,  {Armus} L.,  {Bohlin} R.~C.,  {Kinney} A.~L.,  {Koornneef} J.,
   {Storchi-Bergmann} T.,  2000, \mn@doi [\apj] {10.1086/308692}, \href
  {https://ui.adsabs.harvard.edu/abs/2000ApJ...533..682C} {533, 682}

\bibitem[\protect\citeauthoryear{{Chabrier}}{{Chabrier}}{2003}]{CHA03}
{Chabrier} G.,  2003, \mn@doi [\pasp] {10.1086/376392}, \href
  {https://ui.adsabs.harvard.edu/abs/2003PASP..115..763C} {115, 763}

\bibitem[\protect\citeauthoryear{{Chen}, {Helsby}, {Gauthier}, {Shectman},
  {Thompson}  \& {Tinker}}{{Chen} et~al.}{2010}]{Chen2010}
{Chen} H.-W.,  {Helsby} J.~E.,  {Gauthier} J.-R.,  {Shectman} S.~A.,
  {Thompson} I.~B.,   {Tinker} J.~L.,  2010, \mn@doi [\apj]
  {10.1088/0004-637X/714/2/1521}, \href
  {https://ui.adsabs.harvard.edu/abs/2010ApJ...714.1521C} {714, 1521}

\bibitem[\protect\citeauthoryear{{Chilingarian}, {Melchior}  \&
  {Zolotukhin}}{{Chilingarian} et~al.}{2010}]{CHI10}
{Chilingarian} I.~V.,  {Melchior} A.-L.,   {Zolotukhin} I.~Y.,  2010, \mn@doi
  [\mnras] {10.1111/j.1365-2966.2010.16506.x}, \href
  {https://ui.adsabs.harvard.edu/abs/2010MNRAS.405.1409C} {405, 1409}

\bibitem[\protect\citeauthoryear{{Chiosi} \& {Carraro}}{{Chiosi} \&
  {Carraro}}{2002}]{CC02}
{Chiosi} C.,  {Carraro} G.,  2002, \mn@doi [\mnras]
  {10.1046/j.1365-8711.2002.05590.x}, \href
  {https://ui.adsabs.harvard.edu/abs/2002MNRAS.335..335C} {335, 335}

\bibitem[\protect\citeauthoryear{{Clocchiatti} \& {Wheeler}}{{Clocchiatti} \&
  {Wheeler}}{1997}]{CLO97}
{Clocchiatti} A.,  {Wheeler} J.~C.,  1997, in {Ruiz-Lapuente} P.,  {Canal} R.,
   {Isern} J.,  eds,  NATO Advanced Study Institute (ASI) Series C Vol. 486,
  Thermonuclear Supernovae. p.~863 (\mn@eprint {arXiv} {astro-ph/9601023}),
  \mn@doi{10.1007/978-94-011-5710-0\_53}

\bibitem[\protect\citeauthoryear{{Cole}, {Lacey}, {Baugh}  \& {Frenk}}{{Cole}
  et~al.}{2000}]{COL00}
{Cole} S.,  {Lacey} C.~G.,  {Baugh} C.~M.,   {Frenk} C.~S.,  2000, \mn@doi
  [\mnras] {10.1046/j.1365-8711.2000.03879.x}, \href
  {https://ui.adsabs.harvard.edu/abs/2000MNRAS.319..168C} {319, 168}

\bibitem[\protect\citeauthoryear{{Dawson} et~al.,}{{Dawson}
  et~al.}{2013}]{DAW13}
{Dawson} K.~S.,  et~al., 2013, \mn@doi [\aj] {10.1088/0004-6256/145/1/10},
  \href {https://ui.adsabs.harvard.edu/abs/2013AJ....145...10D} {145, 10}

\bibitem[\protect\citeauthoryear{{Dawson} et~al.,}{{Dawson}
  et~al.}{2016}]{DAW16}
{Dawson} K.~S.,  et~al., 2016, \mn@doi [\aj] {10.3847/0004-6256/151/2/44},
  \href {https://ui.adsabs.harvard.edu/abs/2016AJ....151...44D} {151, 44}

\bibitem[\protect\citeauthoryear{{Driver} \& {Robotham}}{{Driver} \&
  {Robotham}}{2010}]{DR10}
{Driver} S.~P.,  {Robotham} A. S.~G.,  2010, \mn@doi [\mnras]
  {10.1111/j.1365-2966.2010.17028.x}, \href
  {https://ui.adsabs.harvard.edu/abs/2010MNRAS.407.2131D} {407, 2131}

\bibitem[\protect\citeauthoryear{{Dubois}, {Peirani}, {Pichon}, {Devriendt},
  {Gavazzi}, {Welker}  \& {Volonteri}}{{Dubois} et~al.}{2016}]{DUB16}
{Dubois} Y.,  {Peirani} S.,  {Pichon} C.,  {Devriendt} J.,  {Gavazzi} R.,
  {Welker} C.,   {Volonteri} M.,  2016, \mn@doi [\mnras]
  {10.1093/mnras/stw2265}, \href
  {https://ui.adsabs.harvard.edu/abs/2016MNRAS.463.3948D} {463, 3948}

\bibitem[\protect\citeauthoryear{{Eggen}, {Lynden-Bell}  \& {Sandage}}{{Eggen}
  et~al.}{1962}]{ELBS62}
{Eggen} O.~J.,  {Lynden-Bell} D.,   {Sandage} A.~R.,  1962, \mn@doi [\apj]
  {10.1086/147433}, \href
  {https://ui.adsabs.harvard.edu/abs/1962ApJ...136..748E} {136, 748}

\bibitem[\protect\citeauthoryear{{Erb}, {Steidel}, {Shapley}, {Pettini},
  {Reddy}  \& {Adelberger}}{{Erb} et~al.}{2006}]{ERB06}
{Erb} D.~K.,  {Steidel} C.~C.,  {Shapley} A.~E.,  {Pettini} M.,  {Reddy} N.~A.,
    {Adelberger} K.~L.,  2006, \mn@doi [\apj] {10.1086/505341}, \href
  {https://ui.adsabs.harvard.edu/abs/2006ApJ...647..128E} {647, 128}

\bibitem[\protect\citeauthoryear{{Fliri} \& {Trujillo}}{{Fliri} \&
  {Trujillo}}{2016}]{FT16}
{Fliri} J.,  {Trujillo} I.,  2016, \mn@doi [\mnras] {10.1093/mnras/stv2686},
  \href {https://ui.adsabs.harvard.edu/abs/2016MNRAS.456.1359F} {456, 1359}

\bibitem[\protect\citeauthoryear{{Franceschini} et~al.,}{{Franceschini}
  et~al.}{2006}]{FRA06}
{Franceschini} A.,  et~al., 2006, \mn@doi [\aap] {10.1051/0004-6361:20054360},
  \href {https://ui.adsabs.harvard.edu/abs/2006A&A...453..397F} {453, 397}

\bibitem[\protect\citeauthoryear{{Franx}}{{Franx}}{1993}]{FRA93}
{Franx} M.,  1993, \mn@doi [\pasp] {10.1086/133282}, \href
  {https://ui.adsabs.harvard.edu/abs/1993PASP..105.1058F} {105, 1058}

\bibitem[\protect\citeauthoryear{{Gallagher}, {Garnavich}, {Caldwell},
  {Kirshner}, {Jha}, {Li}, {Ganeshalingam}  \& {Filippenko}}{{Gallagher}
  et~al.}{2008}]{GAL08}
{Gallagher} J.~S.,  {Garnavich} P.~M.,  {Caldwell} N.,  {Kirshner} R.~P.,
  {Jha} S.~W.,  {Li} W.,  {Ganeshalingam} M.,   {Filippenko} A.~V.,  2008,
  \mn@doi [\apj] {10.1086/590659}, \href
  {https://ui.adsabs.harvard.edu/abs/2008ApJ...685..752G} {685, 752}

\bibitem[\protect\citeauthoryear{{Gallazzi}, {Charlot}, {Brinchmann}, {White}
  \& {Tremonti}}{{Gallazzi} et~al.}{2005}]{GAL05}
{Gallazzi} A.,  {Charlot} S.,  {Brinchmann} J.,  {White} S. D.~M.,   {Tremonti}
  C.~A.,  2005, \mn@doi [\mnras] {10.1111/j.1365-2966.2005.09321.x}, \href
  {https://ui.adsabs.harvard.edu/abs/2005MNRAS.362...41G} {362, 41}

\bibitem[\protect\citeauthoryear{{Gehrels}}{{Gehrels}}{1986}]{GEH86}
{Gehrels} N.,  1986, \mn@doi [\apj] {10.1086/164079}, \href
  {https://ui.adsabs.harvard.edu/abs/1986ApJ...303..336G} {303, 336}

\bibitem[\protect\citeauthoryear{{Glazebrook}, {Blake}, {Economou}, {Lilly}  \&
  {Colless}}{{Glazebrook} et~al.}{1999}]{GLA99}
{Glazebrook} K.,  {Blake} C.,  {Economou} F.,  {Lilly} S.,   {Colless} M.,
  1999, \mn@doi [\mnras] {10.1046/j.1365-8711.1999.02576.x}, \href
  {https://ui.adsabs.harvard.edu/abs/1999MNRAS.306..843G} {306, 843}

\bibitem[\protect\citeauthoryear{{Graur}, {Bianco}  \& {Modjaz}}{{Graur}
  et~al.}{2015}]{GRA15}
{Graur} O.,  {Bianco} F.~B.,   {Modjaz} M.,  2015, \mn@doi [\mnras]
  {10.1093/mnras/stv713}, \href
  {https://ui.adsabs.harvard.edu/abs/2015MNRAS.450..905G} {450, 905}

\bibitem[\protect\citeauthoryear{{Greggio} \& {Renzini}}{{Greggio} \&
  {Renzini}}{1990}]{GRE90}
{Greggio} L.,  {Renzini} A.,  1990, \mn@doi [\apj] {10.1086/169384}, \href
  {https://ui.adsabs.harvard.edu/abs/1990ApJ...364...35G} {364, 35}

\bibitem[\protect\citeauthoryear{{Groves}, {Heckman}  \& {Kauffmann}}{{Groves}
  et~al.}{2006}]{GRO06}
{Groves} B.~A.,  {Heckman} T.~M.,   {Kauffmann} G.,  2006, \mn@doi [\mnras]
  {10.1111/j.1365-2966.2006.10812.x}, \href
  {https://ui.adsabs.harvard.edu/abs/2006MNRAS.371.1559G} {371, 1559}

\bibitem[\protect\citeauthoryear{{Hakobyan}}{{Hakobyan}}{2008}]{HAK08}
{Hakobyan} A.~A.,  2008, \mn@doi [Astrophysics] {10.1007/s10511-008-0008-3},
  \href {https://ui.adsabs.harvard.edu/abs/2008Ap.....51...69H} {51, 69}

\bibitem[\protect\citeauthoryear{{Hummels} et~al.,}{{Hummels}
  et~al.}{2019}]{Hummels2019}
{Hummels} C.~B.,  et~al., 2019, \mn@doi [\apj] {10.3847/1538-4357/ab378f},
  \href {https://ui.adsabs.harvard.edu/abs/2019ApJ...882..156H} {882, 156}

\bibitem[\protect\citeauthoryear{{Inoue}, {Hirashita}  \& {Kamaya}}{{Inoue}
  et~al.}{2000}]{INO00}
{Inoue} A.~K.,  {Hirashita} H.,   {Kamaya} H.,  2000, \mn@doi [\pasj]
  {10.1093/pasj/52.3.539}, \href
  {https://ui.adsabs.harvard.edu/abs/2000PASJ...52..539I} {52, 539}

\bibitem[\protect\citeauthoryear{{Ivezi{\'c}} et~al.,}{{Ivezi{\'c}}
  et~al.}{2019}]{IVE19}
{Ivezi{\'c}} {\v{Z}}.,  et~al., 2019, \mn@doi [\apj]
  {10.3847/1538-4357/ab042c}, \href
  {https://ui.adsabs.harvard.edu/abs/2019ApJ...873..111I} {873, 111}

\bibitem[\protect\citeauthoryear{{Jha}, {Riess}  \& {Kirshner}}{{Jha}
  et~al.}{2007}]{JHA07}
{Jha} S.,  {Riess} A.~G.,   {Kirshner} R.~P.,  2007, \mn@doi [\apj]
  {10.1086/512054}, \href
  {https://ui.adsabs.harvard.edu/abs/2007ApJ...659..122J} {659, 122}

\bibitem[\protect\citeauthoryear{{Jorgensen}, {Franx}  \&
  {Kjaergaard}}{{Jorgensen} et~al.}{1996}]{JFK96}
{Jorgensen} I.,  {Franx} M.,   {Kjaergaard} P.,  1996, \mn@doi [\mnras]
  {10.1093/mnras/280.1.167}, \href
  {https://ui.adsabs.harvard.edu/abs/1996MNRAS.280..167J} {280, 167}

\bibitem[\protect\citeauthoryear{{Jura}}{{Jura}}{1977}]{JUR77}
{Jura} M.,  1977, \mn@doi [\apj] {10.1086/155085}, \href
  {https://ui.adsabs.harvard.edu/abs/1977ApJ...212..634J} {212, 634}

\bibitem[\protect\citeauthoryear{{Kauffmann} et~al.,}{{Kauffmann}
  et~al.}{2003}]{KAU03}
{Kauffmann} G.,  et~al., 2003, \mn@doi [\mnras]
  {10.1111/j.1365-2966.2003.07154.x}, \href
  {https://ui.adsabs.harvard.edu/abs/2003MNRAS.346.1055K} {346, 1055}

\bibitem[\protect\citeauthoryear{{Kaviraj}}{{Kaviraj}}{2014a}]{KAV14a}
{Kaviraj} S.,  2014a, \mn@doi [\mnras] {10.1093/mnrasl/slt136}, \href
  {https://ui.adsabs.harvard.edu/abs/2014MNRAS.437L..41K} {437, L41}

\bibitem[\protect\citeauthoryear{{Kaviraj}}{{Kaviraj}}{2014b}]{KAV14b}
{Kaviraj} S.,  2014b, \mn@doi [\mnras] {10.1093/mnras/stu338}, \href
  {https://ui.adsabs.harvard.edu/abs/2014MNRAS.440.2944K} {440, 2944}

\bibitem[\protect\citeauthoryear{{Kaviraj} et~al.,}{{Kaviraj}
  et~al.}{2007}]{KAV07}
{Kaviraj} S.,  et~al., 2007, \mn@doi [\apjs] {10.1086/516633}, \href
  {https://ui.adsabs.harvard.edu/abs/2007ApJS..173..619K} {173, 619}

\bibitem[\protect\citeauthoryear{{Kaviraj} et~al.,}{{Kaviraj}
  et~al.}{2008}]{KAV08}
{Kaviraj} S.,  et~al., 2008, \mn@doi [\mnras]
  {10.1111/j.1365-2966.2008.13392.x}, \href
  {https://ui.adsabs.harvard.edu/abs/2008MNRAS.388...67K} {388, 67}

\bibitem[\protect\citeauthoryear{{Kaviraj}, {Peirani}, {Khochfar}, {Silk}  \&
  {Kay}}{{Kaviraj} et~al.}{2009}]{KAV09}
{Kaviraj} S.,  {Peirani} S.,  {Khochfar} S.,  {Silk} J.,   {Kay} S.,  2009,
  \mn@doi [\mnras] {10.1111/j.1365-2966.2009.14403.x}, \href
  {https://ui.adsabs.harvard.edu/abs/2009MNRAS.394.1713K} {394, 1713}

\bibitem[\protect\citeauthoryear{{Kaviraj}, {Tan}, {Ellis}  \&
  {Silk}}{{Kaviraj} et~al.}{2011}]{KAV11}
{Kaviraj} S.,  {Tan} K.-M.,  {Ellis} R.~S.,   {Silk} J.,  2011, \mn@doi
  [\mnras] {10.1111/j.1365-2966.2010.17754.x}, \href
  {https://ui.adsabs.harvard.edu/abs/2011MNRAS.411.2148K} {411, 2148}

\bibitem[\protect\citeauthoryear{{Kaviraj} et~al.,}{{Kaviraj}
  et~al.}{2017}]{KAV17}
{Kaviraj} S.,  et~al., 2017, \mn@doi [\mnras] {10.1093/mnras/stx126}, \href
  {https://ui.adsabs.harvard.edu/abs/2017MNRAS.467.4739K} {467, 4739}

\bibitem[\protect\citeauthoryear{{Kelvin} et~al.,}{{Kelvin}
  et~al.}{2014}]{KEL14b}
{Kelvin} L.~S.,  et~al., 2014, \mn@doi [\mnras] {10.1093/mnras/stu1507}, \href
  {https://ui.adsabs.harvard.edu/abs/2014MNRAS.444.1647K} {444, 1647}

\bibitem[\protect\citeauthoryear{{Kennicutt}}{{Kennicutt}}{1998}]{KEN98}
{Kennicutt} Robert~C. J.,  1998, \mn@doi [\araa]
  {10.1146/annurev.astro.36.1.189}, \href
  {https://ui.adsabs.harvard.edu/abs/1998ARA&A..36..189K} {36, 189}

\bibitem[\protect\citeauthoryear{{Khochfar} \& {Burkert}}{{Khochfar} \&
  {Burkert}}{2003}]{KB03}
{Khochfar} S.,  {Burkert} A.,  2003, \mn@doi [\apjl] {10.1086/379845}, \href
  {https://ui.adsabs.harvard.edu/abs/2003ApJ...597L.117K} {597, L117}

\bibitem[\protect\citeauthoryear{{Kron}}{{Kron}}{1980}]{KRO80}
{Kron} R.~G.,  1980, \mn@doi [\apjs] {10.1086/190669}, \href
  {https://ui.adsabs.harvard.edu/#abs/1980ApJS...43..305K} {43, 305}

\bibitem[\protect\citeauthoryear{{Larson}}{{Larson}}{1974}]{LAR74}
{Larson} R.~B.,  1974, \mn@doi [\mnras] {10.1093/mnras/166.3.585}, \href
  {https://ui.adsabs.harvard.edu/abs/1974MNRAS.166..585L} {166, 585}

\bibitem[\protect\citeauthoryear{{Li}, {Chornock}, {Leaman}, {Filippenko},
  {Poznanski}, {Wang}, {Ganeshalingam}  \& {Mannucci}}{{Li}
  et~al.}{2011}]{LI11b}
{Li} W.,  {Chornock} R.,  {Leaman} J.,  {Filippenko} A.~V.,  {Poznanski} D.,
  {Wang} X.,  {Ganeshalingam} M.,   {Mannucci} F.,  2011, \mn@doi [\mnras]
  {10.1111/j.1365-2966.2011.18162.x}, \href
  {https://ui.adsabs.harvard.edu/abs/2011MNRAS.412.1473L} {412, 1473}

\bibitem[\protect\citeauthoryear{{Lintott} et~al.,}{{Lintott}
  et~al.}{2011}]{LIN11}
{Lintott} C.,  et~al., 2011, \mn@doi [\mnras]
  {10.1111/j.1365-2966.2010.17432.x}, \href
  {https://ui.adsabs.harvard.edu/abs/2011MNRAS.410..166L} {410, 166}

\bibitem[\protect\citeauthoryear{{Martin} et~al.,}{{Martin}
  et~al.}{2005}]{MAR05}
{Martin} D.~C.,  et~al., 2005, \mn@doi [\apjl] {10.1086/426387}, \href
  {https://ui.adsabs.harvard.edu/abs/2005ApJ...619L...1M} {619, L1}

\bibitem[\protect\citeauthoryear{{Martin}, {Kaviraj}, {Devriendt}, {Dubois},
  {Laigle}  \& {Pichon}}{{Martin} et~al.}{2017}]{MAR17}
{Martin} G.,  {Kaviraj} S.,  {Devriendt} J.~E.~G.,  {Dubois} Y.,  {Laigle} C.,
   {Pichon} C.,  2017, \mn@doi [\mnras] {10.1093/mnrasl/slx136}, \href
  {https://ui.adsabs.harvard.edu/abs/2017MNRAS.472L..50M} {472, L50}

\bibitem[\protect\citeauthoryear{{Martin}, {Kaviraj}, {Devriendt}, {Dubois}  \&
  {Pichon}}{{Martin} et~al.}{2018}]{MAR18}
{Martin} G.,  {Kaviraj} S.,  {Devriendt} J.~E.~G.,  {Dubois} Y.,   {Pichon} C.,
   2018, \mn@doi [\mnras] {10.1093/mnras/sty1936}, \href
  {https://ui.adsabs.harvard.edu/abs/2018MNRAS.480.2266M} {480, 2266}

\bibitem[\protect\citeauthoryear{{McGaugh}, {Schombert}  \& {Lelli}}{{McGaugh}
  et~al.}{2017}]{MCG17}
{McGaugh} S.~S.,  {Schombert} J.~M.,   {Lelli} F.,  2017, \mn@doi [\apj]
  {10.3847/1538-4357/aa9790}, \href
  {https://ui.adsabs.harvard.edu/abs/2017ApJ...851...22M} {851, 22}

\bibitem[\protect\citeauthoryear{{Morrissey} et~al.,}{{Morrissey}
  et~al.}{2007}]{MOR07}
{Morrissey} P.,  et~al., 2007, \mn@doi [\apjs] {10.1086/520512}, \href
  {https://ui.adsabs.harvard.edu/abs/2007ApJS..173..682M} {173, 682}

\bibitem[\protect\citeauthoryear{{Noeske} et~al.,}{{Noeske}
  et~al.}{2007}]{NOE07}
{Noeske} K.~G.,  et~al., 2007, \mn@doi [\apjl] {10.1086/517926}, \href
  {https://ui.adsabs.harvard.edu/abs/2007ApJ...660L..43N} {660, L43}

\bibitem[\protect\citeauthoryear{{Noll}, {Burgarella}, {Giovannoli}, {Buat},
  {Marcillac}  \& {Mu{\~n}oz-Mateos}}{{Noll} et~al.}{2009}]{NOL09}
{Noll} S.,  {Burgarella} D.,  {Giovannoli} E.,  {Buat} V.,  {Marcillac} D.,
  {Mu{\~n}oz-Mateos} J.~C.,  2009, \mn@doi [\aap]
  {10.1051/0004-6361/200912497}, \href
  {https://ui.adsabs.harvard.edu/abs/2009A&A...507.1793N} {507, 1793}

\bibitem[\protect\citeauthoryear{{Oparin} \& {Moiseev}}{{Oparin} \&
  {Moiseev}}{2018}]{OPA18}
{Oparin} D.~V.,  {Moiseev} A.~V.,  2018, \mn@doi [Astrophysical Bulletin]
  {10.1134/S1990341318030045}, \href
  {https://ui.adsabs.harvard.edu/abs/2018AstBu..73..298O} {73, 298}

\bibitem[\protect\citeauthoryear{{Peebles}}{{Peebles}}{2002}]{PEE02}
{Peebles} P. J.~E.,  2002, in {Metcalfe} N.,  {Shanks} T.,  eds,  Astronomical
  Society of the Pacific Conference Series Vol. 283, A New Era in Cosmology.
  p.~351 (\mn@eprint {arXiv} {astro-ph/0201015})

\bibitem[\protect\citeauthoryear{{Pillepich} et~al.,}{{Pillepich}
  et~al.}{2018}]{PIL18}
{Pillepich} A.,  et~al., 2018, \mn@doi [\mnras] {10.1093/mnras/stx2656}, \href
  {https://ui.adsabs.harvard.edu/abs/2018MNRAS.473.4077P} {473, 4077}

\bibitem[\protect\citeauthoryear{{Preston}, {Sneden}, {Thompson}, {Shectman}
  \& {Burley}}{{Preston} et~al.}{2006}]{PRE06}
{Preston} G.~W.,  {Sneden} C.,  {Thompson} I.~B.,  {Shectman} S.~A.,   {Burley}
  G.~S.,  2006, \mn@doi [\aj] {10.1086/504425}, \href
  {https://ui.adsabs.harvard.edu/abs/2006AJ....132...85P} {132, 85}

\bibitem[\protect\citeauthoryear{{Richardson}, {Jenkins}, {Wright}  \&
  {Maddox}}{{Richardson} et~al.}{2014}]{R14}
{Richardson} D.,  {Jenkins} Robert~L. I.,  {Wright} J.,   {Maddox} L.,  2014,
  \mn@doi [\aj] {10.1088/0004-6256/147/5/118}, \href
  {https://ui.adsabs.harvard.edu/abs/2014AJ....147..118R} {147, 118}

\bibitem[\protect\citeauthoryear{{Saglia et al.}}{{Saglia et
  al.}}{1997}]{SAG97}
{Saglia et al.} 1997, in {Arnaboldi} M.,  {Da Costa} G.~S.,   {Saha} P.,  eds,
  Astronomical Society of the Pacific Conference Series Vol. 116, The Nature of
  Elliptical Galaxies; 2nd Stromlo Symposium. p.~180

\bibitem[\protect\citeauthoryear{{Sako} et~al.,}{{Sako} et~al.}{2011}]{SAK11}
{Sako} M.,  et~al., 2011, \mn@doi [\apj] {10.1088/0004-637X/738/2/162}, \href
  {https://ui.adsabs.harvard.edu/abs/2011ApJ...738..162S} {738, 162}

\bibitem[\protect\citeauthoryear{{Sako} et~al.,}{{Sako} et~al.}{2018}]{SAK18}
{Sako} M.,  et~al., 2018, \mn@doi [\pasp] {10.1088/1538-3873/aab4e0}, \href
  {https://ui.adsabs.harvard.edu/abs/2018PASP..130f4002S} {130, 064002}

\bibitem[\protect\citeauthoryear{{Salim} et~al.,}{{Salim} et~al.}{2007}]{SAL07}
{Salim} S.,  et~al., 2007, \mn@doi [\apjs] {10.1086/519218}, \href
  {https://ui.adsabs.harvard.edu/abs/2007ApJS..173..267S} {173, 267}

\bibitem[\protect\citeauthoryear{{Salim} et~al.,}{{Salim} et~al.}{2016}]{SAL16}
{Salim} S.,  et~al., 2016, \mn@doi [\apjs] {10.3847/0067-0049/227/1/2}, \href
  {https://ui.adsabs.harvard.edu/abs/2016ApJS..227....2S} {227, 2}

\bibitem[\protect\citeauthoryear{{Salim}, {Boquien}  \& {Lee}}{{Salim}
  et~al.}{2018}]{SAL18}
{Salim} S.,  {Boquien} M.,   {Lee} J.~C.,  2018, \mn@doi [\apj]
  {10.3847/1538-4357/aabf3c}, \href
  {https://ui.adsabs.harvard.edu/abs/2018ApJ...859...11S} {859, 11}

\bibitem[\protect\citeauthoryear{{Schaye} et~al.,}{{Schaye}
  et~al.}{2015}]{SCH15}
{Schaye} J.,  et~al., 2015, \mn@doi [\mnras] {10.1093/mnras/stu2058}, \href
  {https://ui.adsabs.harvard.edu/abs/2015MNRAS.446..521S} {446, 521}

\bibitem[\protect\citeauthoryear{{Schechter}}{{Schechter}}{1976}]{SCH76}
{Schechter} P.,  1976, \mn@doi [\apj] {10.1086/154079}, \href
  {https://ui.adsabs.harvard.edu/abs/1976ApJ...203..297S} {203, 297}

\bibitem[\protect\citeauthoryear{{Schlegel}, {Finkbeiner}  \&
  {Davis}}{{Schlegel} et~al.}{1998}]{SFD98}
{Schlegel} D.~J.,  {Finkbeiner} D.~P.,   {Davis} M.,  1998, \mn@doi [\apj]
  {10.1086/305772}, \href
  {https://ui.adsabs.harvard.edu/abs/1998ApJ...500..525S} {500, 525}

\bibitem[\protect\citeauthoryear{{Schweizer} \& {Seitzer}}{{Schweizer} \&
  {Seitzer}}{1992}]{SS92}
{Schweizer} F.,  {Seitzer} P.,  1992, \mn@doi [\aj] {10.1086/116296}, \href
  {https://ui.adsabs.harvard.edu/abs/1992AJ....104.1039S} {104, 1039}

\bibitem[\protect\citeauthoryear{{Schweizer}, {Seitzer}, {Faber}, {Burstein},
  {Dalle Ore}  \& {Gonzalez}}{{Schweizer} et~al.}{1990}]{SCH90}
{Schweizer} F.,  {Seitzer} P.,  {Faber} S.~M.,  {Burstein} D.,  {Dalle Ore}
  C.~M.,   {Gonzalez} J.~J.,  1990, \mn@doi [\apjl] {10.1086/185868}, \href
  {https://ui.adsabs.harvard.edu/abs/1990ApJ...364L..33S} {364, L33}

\bibitem[\protect\citeauthoryear{Sedgwick, Baldry, James  \& Kelvin}{Sedgwick
  et~al.}{2019a}]{SED19b}
Sedgwick T.~M.,  Baldry I.~K.,  James P.~A.,   Kelvin L.~S.,  2019a, in {IAU
  Symposium 355}: {The Realm of the Low Surface Brightness Universe}.
  (\mn@eprint {arXiv} {1909.04535})

\bibitem[\protect\citeauthoryear{{Sedgwick}, {Baldry}, {James}  \&
  {Kelvin}}{{Sedgwick} et~al.}{2019b}]{SED19a}
{Sedgwick} T.~M.,  {Baldry} I.~K.,  {James} P.~A.,   {Kelvin} L.~S.,  2019b,
  \mn@doi [\mnras] {10.1093/mnras/stz186}, \href
  {https://ui.adsabs.harvard.edu/abs/2019MNRAS.484.5278S} {484, 5278}

\bibitem[\protect\citeauthoryear{{Smartt}}{{Smartt}}{2009}]{SMA09}
{Smartt} S.~J.,  2009, \mn@doi [\araa] {10.1146/annurev-astro-082708-101737},
  \href {https://ui.adsabs.harvard.edu/abs/2009ARA&A..47...63S} {47, 63}

\bibitem[\protect\citeauthoryear{{Speagle}, {Steinhardt}, {Capak}  \&
  {Silverman}}{{Speagle} et~al.}{2014}]{SPE14}
{Speagle} J.~S.,  {Steinhardt} C.~L.,  {Capak} P.~L.,   {Silverman} J.~D.,
  2014, \mn@doi [\apjs] {10.1088/0067-0049/214/2/15}, \href
  {https://ui.adsabs.harvard.edu/abs/2014ApJS..214...15S} {214, 15}

\bibitem[\protect\citeauthoryear{{Thom} et~al.,}{{Thom}
  et~al.}{2012}]{Thom2012}
{Thom} C.,  et~al., 2012, \mn@doi [\apjl] {10.1088/2041-8205/758/2/L41}, \href
  {https://ui.adsabs.harvard.edu/abs/2012ApJ...758L..41T} {758, L41}

\bibitem[\protect\citeauthoryear{{Tumlinson}, {Peeples}  \& {Werk}}{{Tumlinson}
  et~al.}{2017}]{Tumlinson2017}
{Tumlinson} J.,  {Peeples} M.~S.,   {Werk} J.~K.,  2017, \mn@doi [\araa]
  {10.1146/annurev-astro-091916-055240}, \href
  {https://ui.adsabs.harvard.edu/abs/2017ARA&A..55..389T} {55, 389}

\bibitem[\protect\citeauthoryear{{Turner} et~al.,}{{Turner}
  et~al.}{2021}]{TUR21}
{Turner} S.,  et~al., 2021, \mn@doi [\mnras] {10.1093/mnras/stab653}, \href
  {https://ui.adsabs.harvard.edu/abs/2021MNRAS.503.3010T} {503, 3010}

\bibitem[\protect\citeauthoryear{{Van Dokkum}, {Franx}, {Fabricant},
  {Illingworth}  \& {Kelson}}{{Van Dokkum} et~al.}{2000}]{VD00}
{Van Dokkum} P.~G.,  {Franx} M.,  {Fabricant} D.,  {Illingworth} G.~D.,
  {Kelson} D.~D.,  2000, \mn@doi [\apj] {10.1086/309402}, \href
  {https://ui.adsabs.harvard.edu/abs/2000ApJ...541...95V} {541, 95}

\bibitem[\protect\citeauthoryear{{Veilleux} \& {Osterbrock}}{{Veilleux} \&
  {Osterbrock}}{1987}]{VO87}
{Veilleux} S.,  {Osterbrock} D.~E.,  1987, \mn@doi [\apjs] {10.1086/191166},
  \href {https://ui.adsabs.harvard.edu/abs/1987ApJS...63..295V} {63, 295}

\bibitem[\protect\citeauthoryear{{Vulcani} et~al.,}{{Vulcani}
  et~al.}{2011}]{VUL11}
{Vulcani} B.,  et~al., 2011, \mn@doi [\mnras]
  {10.1111/j.1365-2966.2010.17904.x}, \href
  {https://ui.adsabs.harvard.edu/abs/2011MNRAS.412..246V} {412, 246}

\bibitem[\protect\citeauthoryear{{Yi}}{{Yi}}{2003}]{YI03}
{Yi} S.~K.,  2003, \mn@doi [\apj] {10.1086/344640}, \href
  {https://ui.adsabs.harvard.edu/abs/2003ApJ...582..202Y} {582, 202}

\bibitem[\protect\citeauthoryear{{York} et~al.,}{{York} et~al.}{2000}]{YOR00}
{York} D.~G.,  et~al., 2000, \mn@doi [\aj] {10.1086/301513}, \href
  {https://ui.adsabs.harvard.edu/abs/2000AJ....120.1579Y} {120, 1579}

\makeatother
\end{thebibliography}
\appendix

\label{lastpage}
\end{document}